\documentclass[aps, prd, amsmath, floats, floatfix, twocolumn, nofootinbib, showpacs]{revtex4-1}
\usepackage{graphicx}
\usepackage{color}

\newcommand{\be}{\begin{eqnarray}}
\newcommand{\ee}{\end{eqnarray}}

\begin{document}

\title{Distinguishing black holes and wormholes with orbiting hot spots}

\author{Zilong Li}

\author{Cosimo Bambi}
\email[Corresponding author: ]{bambi@fudan.edu.cn}

\affiliation{Center for Field Theory and Particle Physics \& Department of Physics, Fudan University, 200433 Shanghai, China}

\date{\today}

\begin{abstract}
The supermassive black hole candidates at the center of every normal galaxy 
might be wormholes created in the early Universe and connecting either two 
different regions of our Universe or two different universes in a Multiverse model. 
Indeed, the origin of these supermassive objects is not well understood, 
topological non-trivial structures like wormholes are allowed both in general
relativity and in alternative theories of gravity, and current observations cannot 
rule out such a possibility. In a few years, the VLTI instrument GRAVITY will have 
the capability to image blobs of plasma orbiting near the innermost stable circular 
orbit of SgrA$^*$, the supermassive black hole candidate in the Milky Way. The
secondary image of a hot spot orbiting around a wormhole is substantially
different from that of a hot spot around a black hole, because the photon
capture sphere of the wormhole is much smaller. The radius of the photon 
capture sphere is independent of the hot spot model, and therefore its possible 
detection, which is observationally challenging but not out of reach, can 
unambiguously test if the center of our Galaxy harbors a wormhole rather than 
a black hole.
\end{abstract}


\maketitle


\section{Introduction}

The Einstein equations are local equations relating the geometry of the spacetime 
to its matter content. There is no information about the spacetime topology. Even if 
it is far from our common sense, we cannot exclude that our Universe has a non-trivial 
topology or that it contains topologically non-trivial structures. In this spirit, there is a 
rich literature on wormholes (WHs); that is, short-cuts connecting two different regions 
of our Universe or two different universes in Multiverse theories~\cite{rev-wh}. WH 
spacetimes are allowed even in alternative theories of gravity. Since it is required 
a change of topology, WHs may not be created in the Universe today, but a number 
of mechanisms may have worked in the early Universe~\cite{wh}. Primordial WHs 
may have survived till today and live somewhere in the Universe. In particular, 
they have been proposed as candidates for the supermassive objects that are seen 
at the center of every normal galaxy~\cite{smbh}. These objects are usually supposed 
to be Kerr black holes (BHs) with masses $M \sim 10^5 - 10^9$~$M_\odot$, but their 
actual nature is not known.

In the case of the center of our Galaxy, the central supermassive object has a mass 
$M \approx 4 \cdot 10^6$~$M_\odot$~\cite{mass}. An upper bound on its radius 
can be inferred from the closest distance approached by the orbiting stars. Current 
data put this bound at about 45~AU, which corresponds to $\sim 600$~Schwarzschild 
radii~\cite{radius}. Such estimates of the mass and radius can exclude the possibility 
that the central object is actually a cluster of neutron stars, because the cluster lifetime 
would only be $\sim 10^5$~yrs, which is much shorter than the age of this 
system~\cite{radius, maoz}. The non-observations of thermal radiation emitted by the 
possible surface of this object may also be interpreted as an indication for the presence
of an event/apparent horizon~\cite{eh}. This body of evidence strongly supports the 
conclusion that the supermassive object at the center of the Galaxy is a supermassive
BH. Such a conclusion is naturally extended to all the supermassive objects in
galactic nuclei. However, there is no real indication that the spacetime geometry around 
these bodies is described by the Kerr solution~\cite{tests} (for a review, see e.g.~\cite{review}).
There are also open questions about their formation and growth. Any competitive 
model must be able to explain how they were able to become so heavy in a very 
short time, as we know of BH candidates with mass $M \sim 10^9$~$M_\odot$ at 
redshift $z \gtrsim 6$, i.e. just about 100~million years after the Big Bang~\cite{super}. 
While there are potentially many possibilities, we do not know which one is 
correct~\cite{mv}. Scenarios in which the seeds of these supermassive objects 
are a relic of the very early Universe are also possible~\cite{dolgov}, and in this 
context there is room for WHs too~\cite{wh}.

While of exotic nature, at least some kinds of primordial WHs can be viable candidates 
to explain the supermassive objects at the centers of galaxies. These objects have no 
solid surface, and therefore they may mimic the presence of an event horizon. They 
would have been produced in the early Universe and grown during inflation, which 
could explain their presence even at very high redshift. They have a positive effective mass, 
and therefore a WH can swallow material from an accretion disk, and two WHs can merge 
when their host galaxies merge. All these processes make the effective mass of the 
WH increase, as the new material is trapped in the WH throat and, even if the latter is 
long, there is no or small spread of the gravitational strength lines. WHs can behave 
in the same way as BHs, even if the physics inside the WH throat can be very 
different from that inside the BH event horizon. For these reasons, WHs can be 
as good as BHs to explain the observed correlations between the supermassive 
BH candidates and the host galaxies~\cite{hostg}. Another important issue is the
formation of relativistic jets, which are commonly observed from BH candidates.
At present, we do not know the actual mechanism responsible for the formation 
of these jets, and there are several models in the literature. If the jets are powered 
by the rotational energy of the compact object, the Blandford-Znajek process~\cite{bz-m}, 
the WHs discussed in this paper cannot have a jet, as they are non-rotating. 
Fast-rotating WHs seem to be already ruled out by current observations~\cite{wh1}.
However, all the other mechanisms in which jets are powered by the rotational 
energy of the disk or by the mass accretion rate (see e.g.~\cite{jets}) can perfectly 
work around a WH without any modification. Current observations cannot yet
provide an answer, and the topic is very controversial~\cite{jets2}.

On the basis of these considerations, it seems that WHs can be serious
candidates to explain the supermassive objects in galactic nuclei.
It is therefore a natural question 
to wonder whether astrophysical observations can distinguish Kerr BHs and WHs and 
thus test the WH scenario. Despite the clear difference between BHs and WHs, 
current observations cannot rule out the possibility that the supermassive objects
in galactic nuclei are WHs instead of BHs. In Ref.~\cite{wh1}, one of us considered a 
particular family of traversable WHs and studied the possibility of distinguishing BHs 
and WHs from the analysis of the iron K$\alpha$ line. The latter is a very narrow 
emission line at about 6.4~keV, but the one observed in the spectra of supermassive
BH candidates is broad and skewed, as a result of special and general relativistic 
effects. In Ref.~\cite{wh1}, it was found that the iron line profile produced in the 
accretion disk of a non-rotating WH looks like the one emitted from a disk around a 
Kerr BH with spin parameter $a_* \approx 0.8$. More in general, WHs with spin 
parameter $a_* \lesssim 0.02$ may be interpreted as Kerr BHs with a spin parameter 
in the range $a_* \approx 0.8 - 1.0$, while WHs with spin parameter larger than 0.02
have substantially different iron lines and can be ruled out because they are inconsistent 
with observations. The constraints found in Ref.~\cite{wh1} and the 
fact that accretion and merger processes should spin the WH up may have two 
possible explanations: $i)$ rotating WHs are unstable and therefore, even if a WH
gets angular momentum, the latter is quickly lost, for instance via the emission of gravitational 
waves, or $ii)$ fast-rotating WHs different from the ones considered in Ref.~\cite{wh1} 
are consistent with current observations; unlike Kerr BHs in general relativity, 
here there is no uniqueness theorem, so the conclusions of Ref.~\cite{wh1} cannot 
be definitive.

The possibility of observationally testing the idea that the supermassive objects
in galactic nuclei are WHs was further discussed in Ref.~\cite{wh2}, where it was
pointed out that it would be relatively easy to figure out if SgrA$^*$ is a WH or
a BH from the observation of its ``shadow''. The latter is a dark area over a brighter
background seen by a distant observer if the compact object is surrounded by 
optically thin emitting material and corresponds to the apparent photon capture 
sphere~\cite{falcke}. While the exact shape and size of the shadow of Kerr and non-Kerr 
BHs is extremely similar~\cite{shadow}, and at present it is not completely clear 
if its future detection can constrain possible deviations from the Kerr 
solution, the shadow of a WH is much smaller than that of a BH, and it is
therefore distinguishable even without an accurate detection and with all
the systematic effects that significantly tangle the job~\cite{wh2}.

In the present paper, we further extend the study of Refs.~\cite{wh1,wh2} and
we investigate the possibility of testing if SgrA$^*$ is a WH by observing a hot 
blob of plasma orbiting near the innermost stable circular orbit (ISCO). Within a few 
years, NIR observations with GRAVITY will have the capability to directly 
image hot spots orbiting near the ISCO of SgrA$^*$~\cite{grav,grav2}, and open a
new window to test the actual nature of this object. These data are supposed to 
come out before the first detection of the shadow. Because of the dramatically 
different sizes of the photon capture spheres of WHs and BHs, we find that the 
possible detection of the secondary image of a hot spot orbiting close to the 
compact object can unambiguously test the possibility that SgrA$^*$ is a WH rather 
than a BH. That will require very good data with a high signal-to-noise ratio, but it is 
not out of reach. Specific differences may also be present in the hot spot light curve 
and in the hot spot centroid track, but their features are more model-dependent: our 
hot spot model is too simple to conclude that their observation in real data can 
distinguish a WH from a BH, and further investigation based on more sophisticated 
and realistic models would be necessary. With the current results, it seems more 
likely that light curves and hot spot centroid tracks cannot test the WH scenarios.

\section{Hot spot model}

SgrA$^*$ exhibits powerful flares in the X-ray, NIR, and sub-mm bands~\cite{flares}.
During a flare, the flux increases by up to a factor of 10. A flare typically lasts 1-3~hours 
and the rate is a few events per day. These flares seem to show a quasi-periodic
substructure on a time scale ranging from 13 to about 30~minutes. Several mechanisms 
have been proposed, such as the heating of electrons in a jet~\cite{m1}, Rossby 
wave instability in the disk~\cite{m2}, the adiabatic expansion of a blob of plasma~\cite{m3}, 
and blobs of plasma orbiting at the ISCO of 
SgrA$^*$~\cite{m4}. At least some authors have claimed that current observations 
favor the model of the hot spot near the ISCO~\cite{trippe}. Such a scenario is also 
supported by some general relativistic magneto-hydrodynamic simulations of 
accretion flows onto BHs that show that temporary clumps of matter may 
be common in the region near the ISCO~\cite{sim}. Within a few years, the GRAVITY 
instrument for the ESO Very Large Telescope Interferometer (VLTI) will have the 
capability to image blobs of plasma orbiting around SgrA$^*$ with an angular resolution 
of about 10~$\mu$as and a time resolution of about 1~minute~\cite{grav,grav2}, 
and it will thus be possible to test the hot spot model.

For a Kerr BH with a mass $M = 4 \cdot 10^6$~$M_\odot$, the ISCO period 
ranges from about 30~minutes ($a_* = 0$) to 4~minutes ($a_* = 1$ and corotating 
orbit). The observed period of the quasi-periodic substructure of the flares of SgrA$^*$
ranges from 13 to about 30 minutes. If the hot spot model is correct, this means that 
the radius of the orbit of the hot spot may vary and be larger than that of the 
ISCO. The shortest period ever measured is $13 \pm 2$~minutes. If we assume 
that the latter corresponds to the ISCO period, or at least that it is very close to it,
one finds $a_* = 0.70 \pm 0.11$ for $M = 3.6 \cdot 10^6$~$M_\odot$~\cite{trippe}.
In Ref.~\cite{asc}, the authors claimed the presence of a quasi-periodic substructure 
with a period of 5~minutes, which was interpreted as an indication that SgrA$^*$ 
is rotating very fast, with a spin parameter close to 1. However, the analysis of the 
same data sets in~\cite{bel} did not find such a short period substructure.

In the present work, we will consider the simplest hot spot model; that is, a single 
region of isotropic and monochromatic emission following a geodesic trajectory. 
Located on the equatorial plane, this hot spot is modeled as an optically thick 
emitting disk of finite radius. The local specific intensity of the radiation is chosen 
to have a Gaussian distribution in the local Cartesian space
\be
\hspace{-0.3cm}
I_{\rm em}(\nu_{\rm em},x) \sim \delta(\nu_{\rm em} - \nu_\star)
\exp\bigg[-\frac{|\textbf{\~{x}}-\textbf{\~{x}}_{\rm spot}(t)|^2}
{2R^2_{\rm spot}}\bigg] \, ,
\label{emissivity}
\ee
where $\nu_{\rm em}$ is the photon frequency measured in the rest-frame of the 
emitter, while $\nu_\star$ is the emission frequency of this monochromatic source. 
The spatial position 3-vector $\textbf{\~{x}}$ is given in pseudo-Cartesian coordinates.
Outside a distance of $4R_{\rm spot}$ from the guiding geodesic trajectory 
$\textbf{\~{x}}_{\rm spot}$, there is no emission. Plausible values are $R_{\rm spot} = 
0.1 - 1.0 M$, but it depends on the orbital radius, and it is important to check that no
point of the hot spot exceeds the speed of light. The specific intensity of 
the radiation measured by the distant observer is given by
\be
I_{\rm obs}(\nu_{\rm obs},t_{\rm obs}) 
= g^3 I_{\rm em} (\nu_{\rm em},t_{\rm obs}) \, ,
\ee
where $g$ is the redshift factor
\be
g=\frac{E_{\rm obs}}{E_{\rm em}} = \frac{\nu_{\rm obs}}{\nu_{\rm em}} =
\frac{k_\alpha u^\alpha _{\rm obs}}{k_\beta u^\beta _{\rm em}} \, ,
\ee
$k^\alpha$ is the 4-momentum of the photon, $u_{\rm obs}^\alpha = (-1, 0, 0, 0)$ 
is the 4-velocity of the distant observer, and $u_{\rm em}^\alpha = (u_{\rm em}^t, 0, 0, 
\Omega u_{\rm em}^t)$ is the 4-velocity of the emitter. $\Omega$ is the Keplerian
angular frequency of a test-particle at the emission radius $r_{\rm e}$. 
$I_{\rm obs}(\nu_{\rm obs})/\nu^3_{\rm obs} = I_{\rm em}(\nu_{\rm em})/\nu^3_{\rm em}$ 
follows from the Liouville theorem. The hot spot emission is assumed to be 
monochromatic and isotropic, with a Gaussian intensity, as shown in Eq.~(\ref{emissivity}). 
Using the normalization condition $g_{\mu\nu}u^\mu_{\rm em}u^\nu_{\rm em} = -1$, 
one finds
\be
u_{\rm em}^t = -\frac{1}{\sqrt{-g_{tt}-2g_{t\phi}
\Omega-g_{\phi\phi}\Omega^2}} \, ,
\ee
and therefore,
\be\label{eq-g}
g = \frac{\sqrt{-g_{tt}-2g_{t\phi}\Omega-g_{\phi\phi}
\Omega^2}}{1+\lambda\Omega} \, , \label{redshift} 
\ee
where $\lambda = k_\phi/k_t$ is a constant of the motion along the photon path. 
Doppler boosting and gravitational redshift are entirely encoded in the redshift factor 
$g$. The effect of light bending is included by the raytracing calculation.

The observer's sky is divided into a number of small elements, and the ray-tracing 
procedure provides the observed time-dependent flux density from each element.
By integrating the observed specific intensity over the solid angle subtended by the 
image of the hot spot on the observer's sky, we obtain the observed flux
\be
F(\nu_{\rm obs},t_{\rm obs}) &=& 
\int I_{\rm obs}(\nu_{\rm obs},t_{\rm obs}) \, \mathrm{d}\Omega_{\rm obs} = 
\nonumber\\
&=& \int g^3 I_{\rm em}(\nu_{\rm em},t_{\rm obs}) 
\, \mathrm{d}\Omega_{\rm obs} \, . 
\ee
If we integrate over the frequency range of the radiation, we get the observed 
luminosity, or light curve, of the hot spot
\be
L(t_{\rm obs}) = \int F(\nu_{\rm obs},t_{\rm obs}) \, \mathrm{d}\nu_{\rm obs} \, . 
\ee
A more detailed description of the calculation procedure can be found, for 
instance, in Ref.~\cite{spot}. In the present paper, we normalize 
the light curves by dividing the observed luminosity $L(t_{\rm obs})$ by the
corresponding maximum, since only the shape of the light curve 
can be used to determine the parameters of the model.
Such a time-dependent emission signal can be added 
to a background intensity coming from the inner region of the steady state 
accretion disk. By definition, the hot spot will have a higher density and/or higher 
temperature, and thus a higher emissivity, than the background accretion disk, 
adding a small modulation to the total flux.

\section{Testing the wormhole scenario}

In the calculations of the electromagnetic radiation emitted by a hot spot, the
spacetime metric determines the exact photon propagation from the hot spot
to the distant observer, the redshift factor $g$ in Eq.~(\ref{eq-g}), and the value
of the ISCO radius. There are many kinds of WHs proposed in the literature, 
but not all are viable supermassive BH candidates. For instance, some WHs
have vanishing or negative effective gravitational mass. In the present work,
we adopt the same asymptotically-flat non-rotating traversable WH solution
discussed in~\cite{wh1,wh2}, whose line element reads~\cite{metric}
\be\label{eq-metric}
ds^2 = e^{2\Phi(r)} dt^2 - \frac{dr^2}{1 - \frac{b(r)}{r}} - r^2 d\theta^2 
- r^2 \sin^2\theta d\phi^2 \, ,
\ee
where $\Phi(r)$ and $b(r)$ are, respectively, the redshift and the shape functions. A 
common choice is $\Phi(r) = - r_0/r$, where $r_0$ is the WH throat radius and sets 
the scales of the system. $r_0$ is interpreted as the mass of the object in the 
Newtonian limit and in what follows it will be indicated with $M$, just to use the 
same notation in the WH and BH cases. The shape function can be assumed
to be of the form
\be
b(r) = \frac{M^\gamma}{r^{\gamma - 1}} \, ,
\ee
where $\gamma$ is a constant. In this paper, we consider the case $\gamma=1$,
but the observational signature that distinguishes WHs and BHs is independent
of the value of $\gamma$.

Let us now compare the features of the electromagnetic radiation emitted by a 
blob of plasma orbiting around a WH and a BH. Since the hot spot orbital frequency 
is the simplest parameter to measure, we want to compare the following three 
cases:
\begin{enumerate}
\item A hot spot orbiting the traversable WH in Eq.~(\ref{eq-metric}) at some 
radius $r_{\rm WH}$. We indicate its angular frequency with $\Omega_{\rm WH}$.
\item A hot spot orbiting at the ISCO radius of a Kerr BH with spin parameter
such that its Keplerian orbital frequency is $\Omega_{\rm ISCO} = \Omega_{\rm WH}$.
\item A hot spot orbiting a Kerr BH with spin parameter $a_* = 0.99$ at the
equatorial circular orbit with radius $r_{\rm BH}$, whose Keplerian orbital
frequency is $\Omega_{\rm BH}=\Omega_{\rm ISCO} = \Omega_{\rm WH}$.
\end{enumerate}

\begin{figure*}
\begin{center}
\includegraphics[type=pdf,ext=.pdf,read=.pdf,width=7cm]{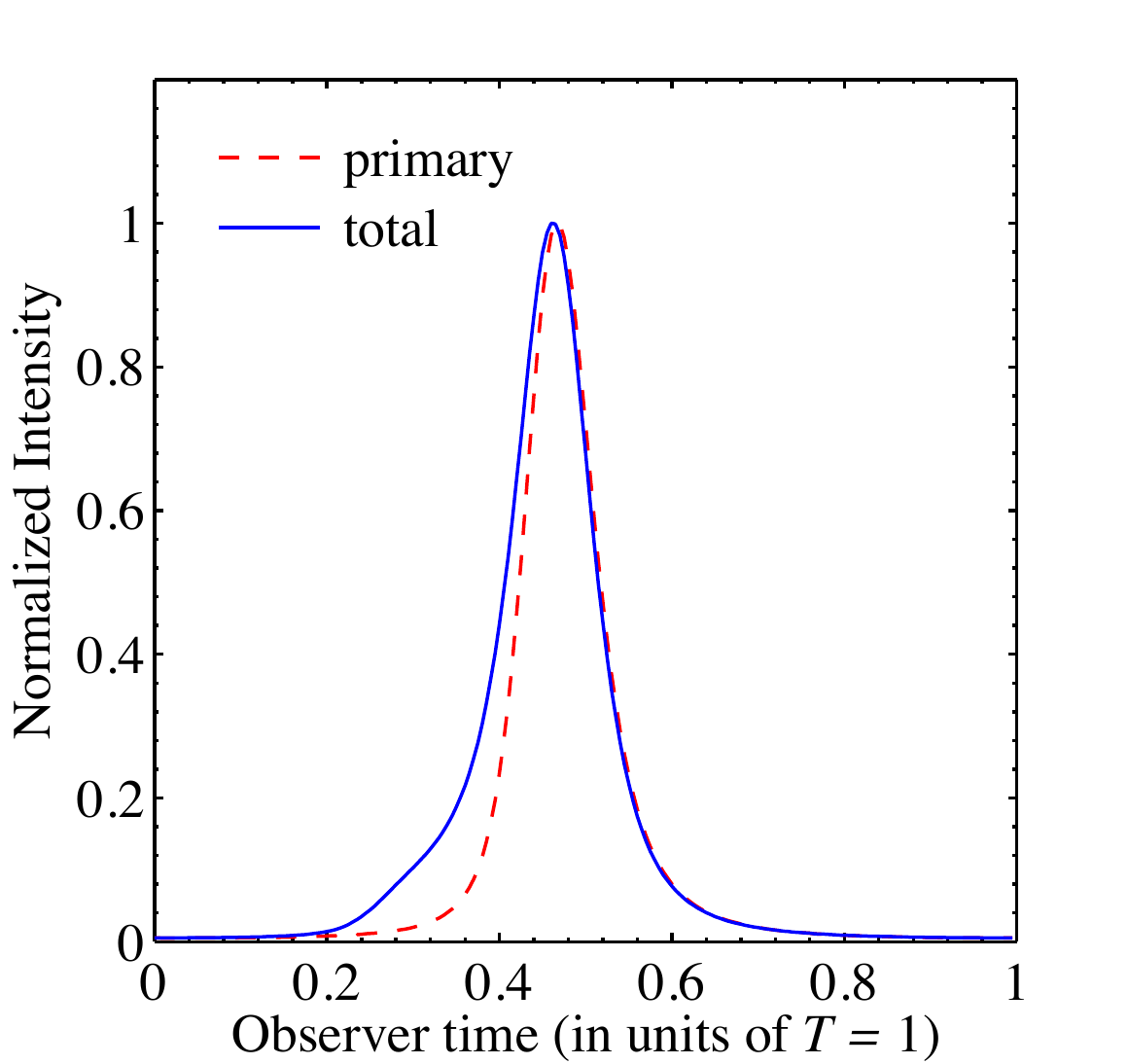} \put(-60,150){\large WH} \hspace{0.5cm}
\includegraphics[type=pdf,ext=.pdf,read=.pdf,width=7.3cm]{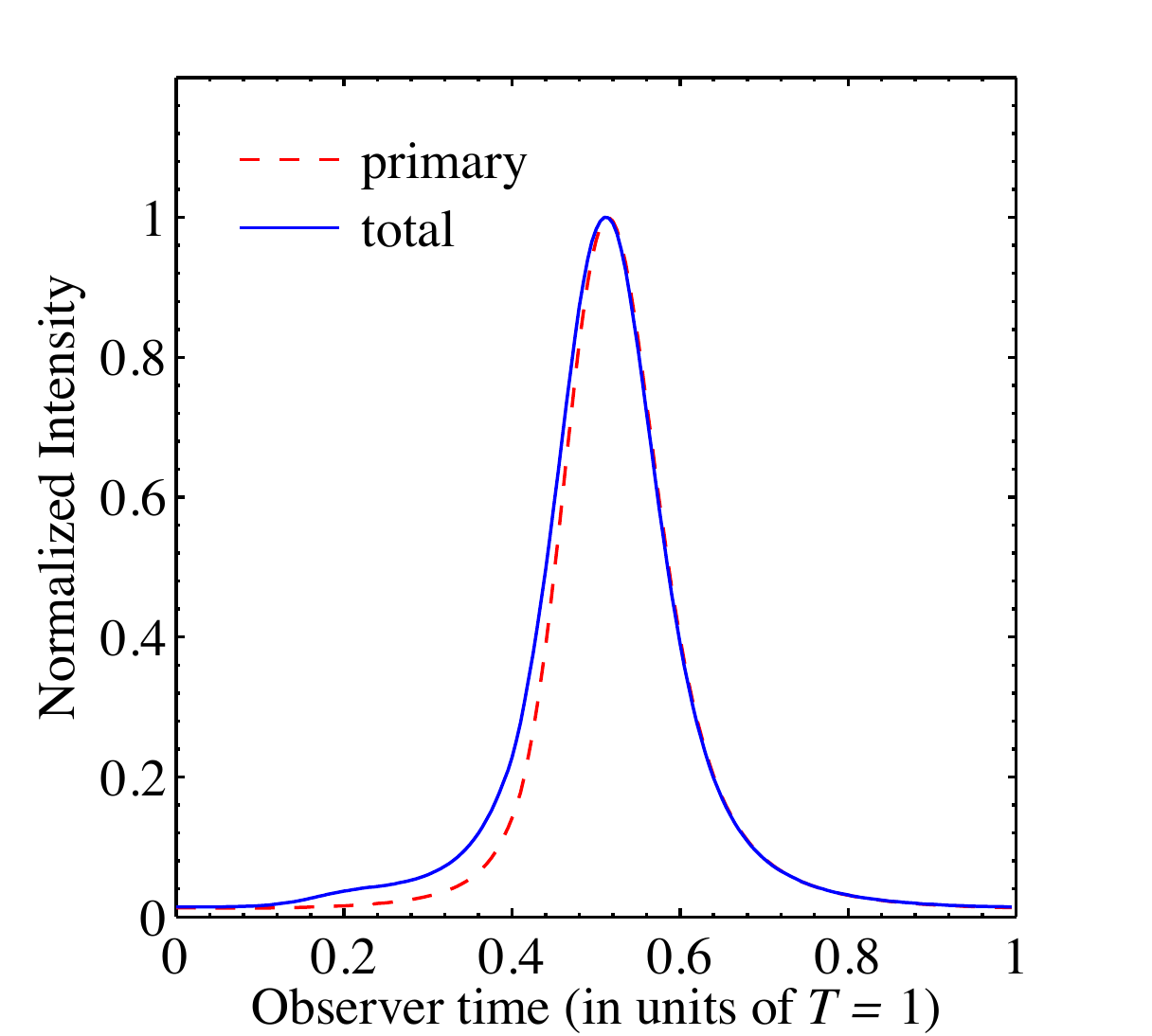} \put(-63,150){\large WH} \\ \vspace{0.5cm}
\includegraphics[type=pdf,ext=.pdf,read=.pdf,width=7cm]{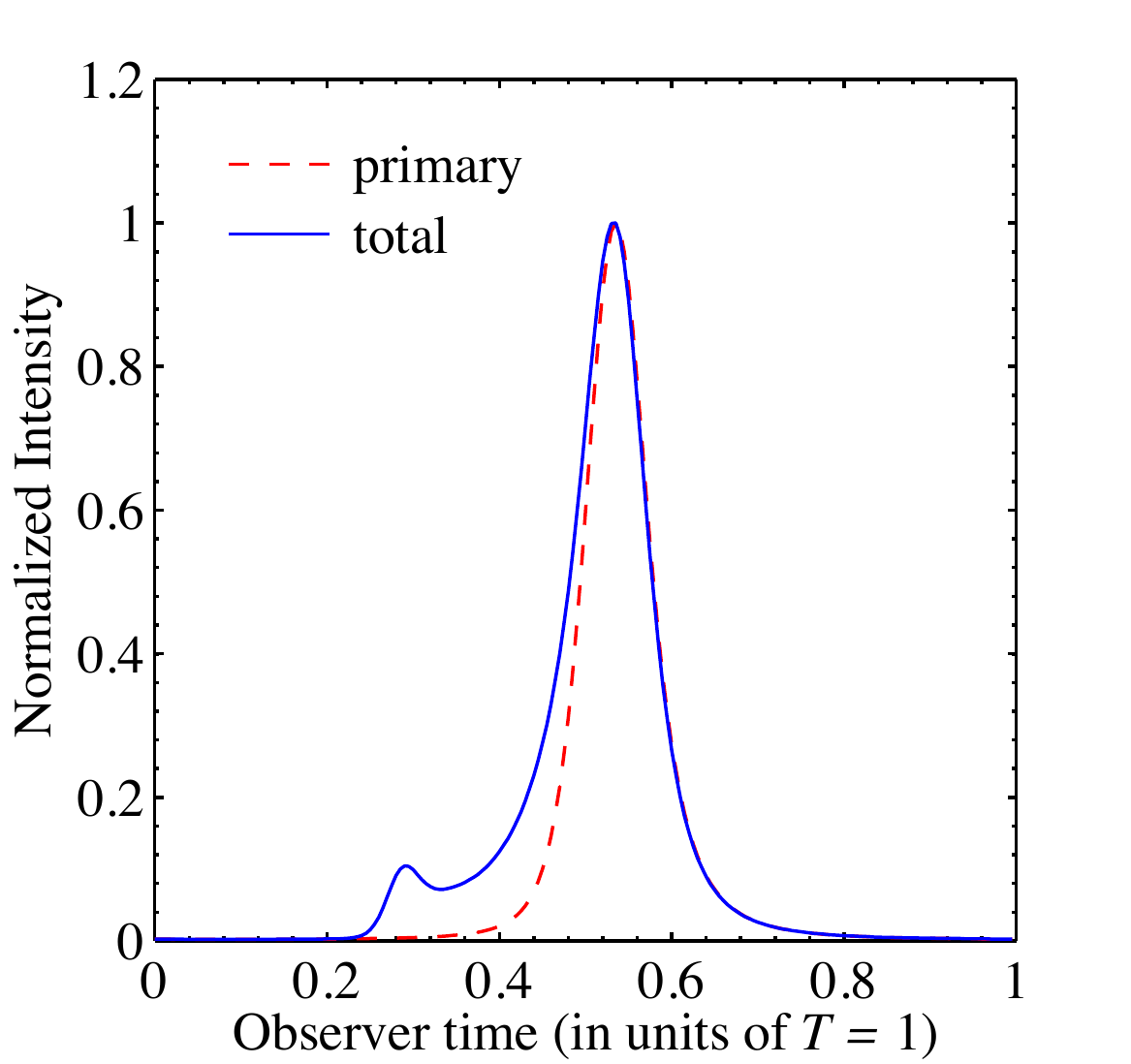} \put(-60,150){\large BH} \hspace{0.5cm}
\includegraphics[type=pdf,ext=.pdf,read=.pdf,width=7.3cm]{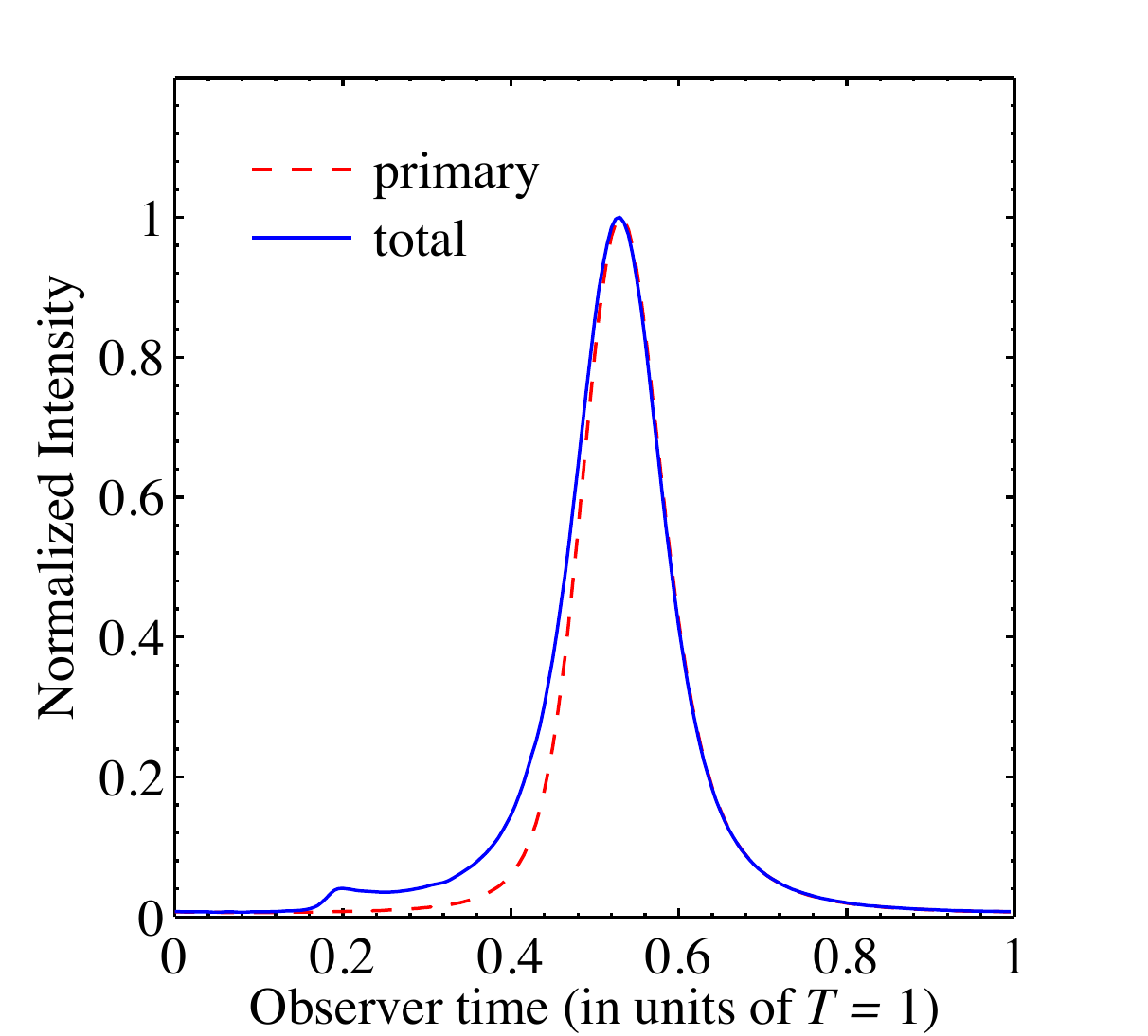} \put(-63,150){\large BH} \\ \vspace{0.5cm}
\includegraphics[type=pdf,ext=.pdf,read=.pdf,width=7cm]{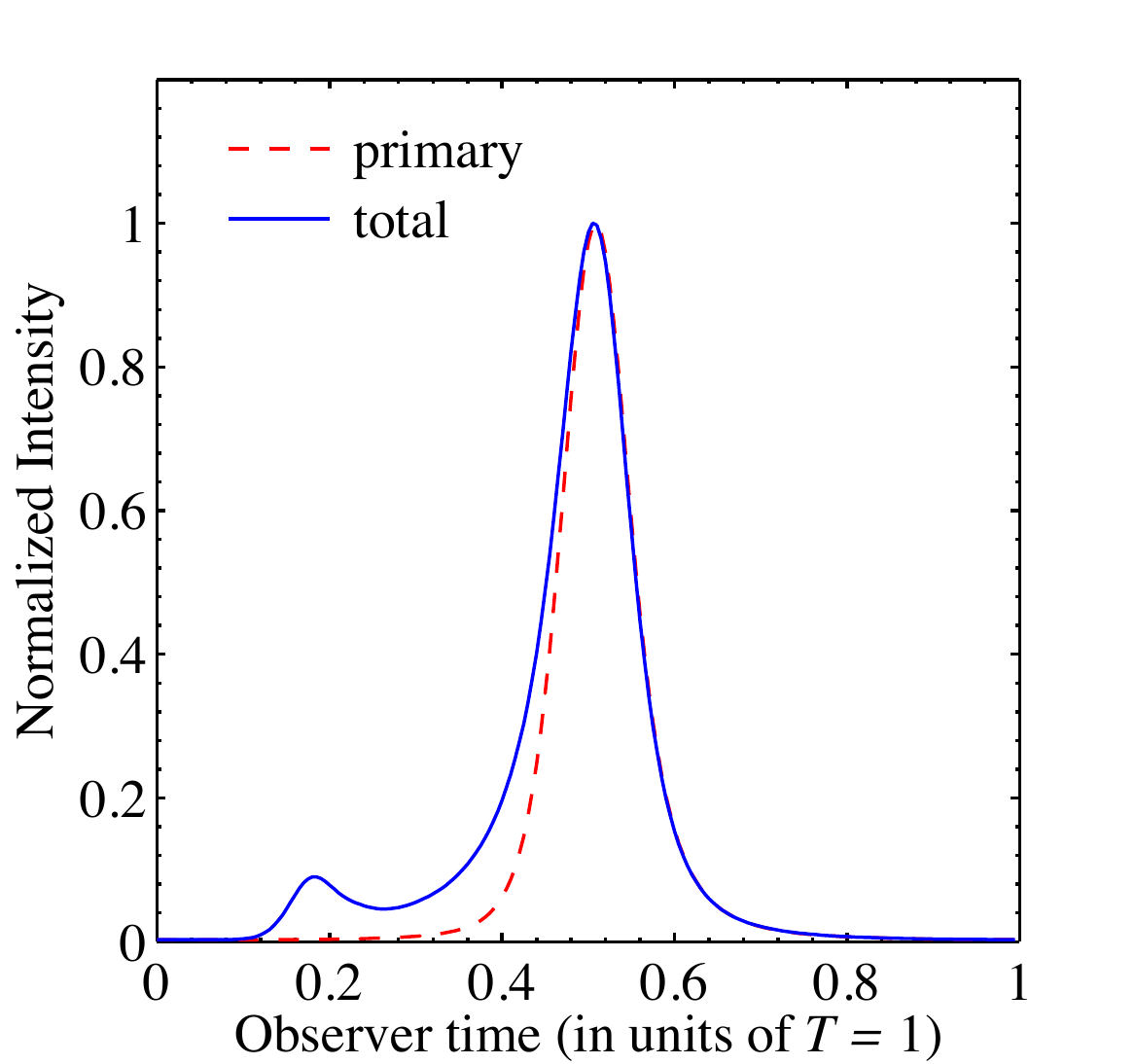} \put(-60,150){\large BH} \hspace{0.5cm}
\includegraphics[type=pdf,ext=.pdf,read=.pdf,width=7.2cm]{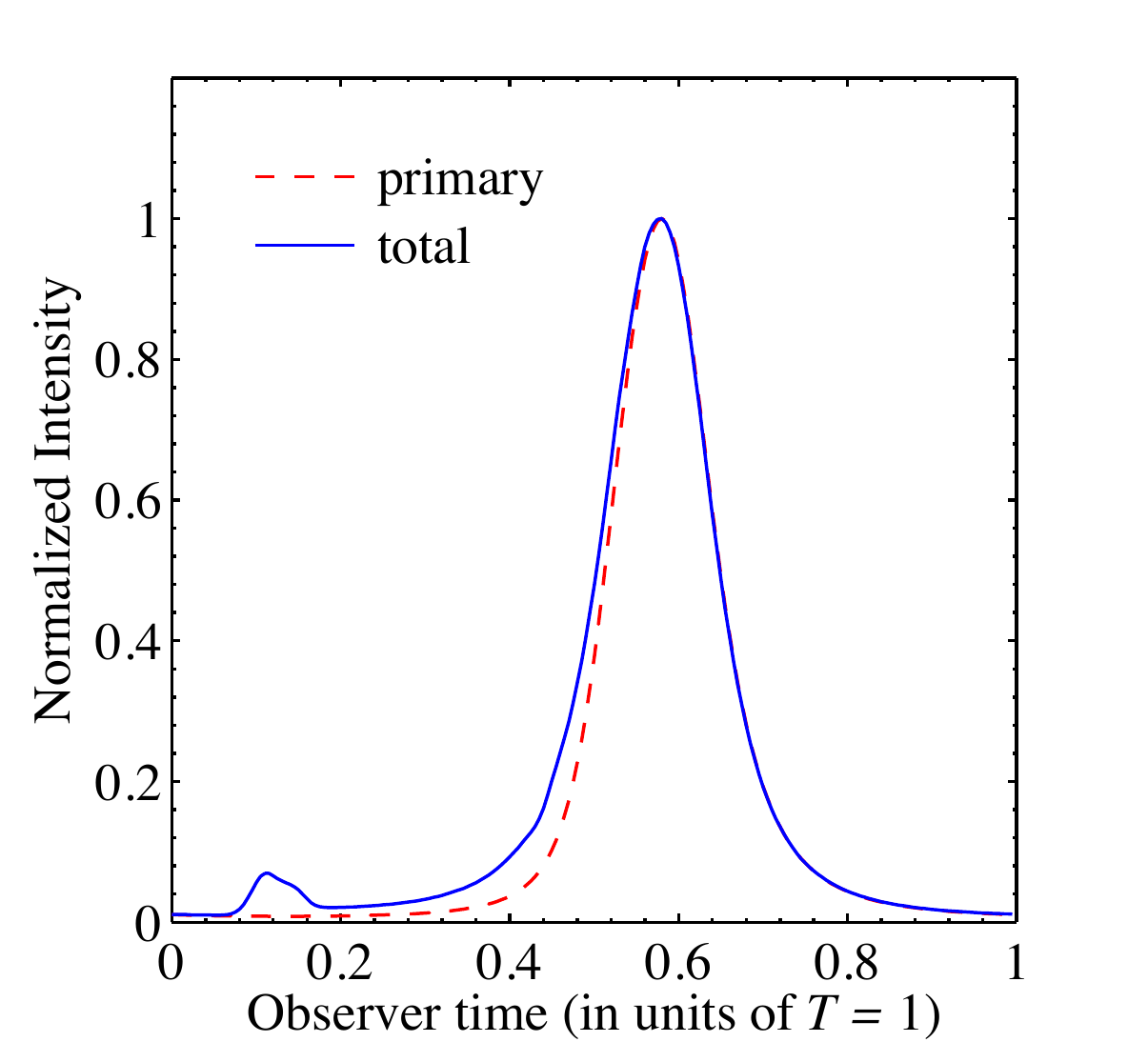} \put(-63,150){\large BH}
\end{center}
\caption{Top panels: total light curves and primary image light curves of a hot spot 
orbiting around a WH at the ISCO $r_{\rm WH}=2\, M$ (left panel) and at the radius 
$r_{\rm WH}=3\, M$ (right panel). The viewing angle of the observer is $i=60^\circ$ 
and the hot spot size is $R_{\rm spot} = 0.15 \, M$. Central panels: as in the top 
panels for a hot spot
orbiting the ISCO of a Kerr BH with spin parameter $a_*=0.883911$ (left panel) 
and $a_*=0.673917$ (right panel); the value of the spin has been chosen to have
an orbital frequency equal, respectively, to that of a hot spot orbiting a WH at 
the ISCO and at the radius $r_{\rm WH}=3\, M$. Bottom panels: as in the top and central 
panels for a hot spot orbiting a Kerr BH with spin parameter $a_*=0.99$ at the
radius with the same Keplerian orbital frequency as that of a hot spot orbiting
a WH at the ISCO (left panel) and at the radius $r_{\rm WH}=3\, M$ (right panel). 
See the text for more details.}
\label{f1}
\end{figure*}

\begin{figure*}
\begin{center}
\includegraphics[type=pdf,ext=.pdf,read=.pdf,width=6.5cm]{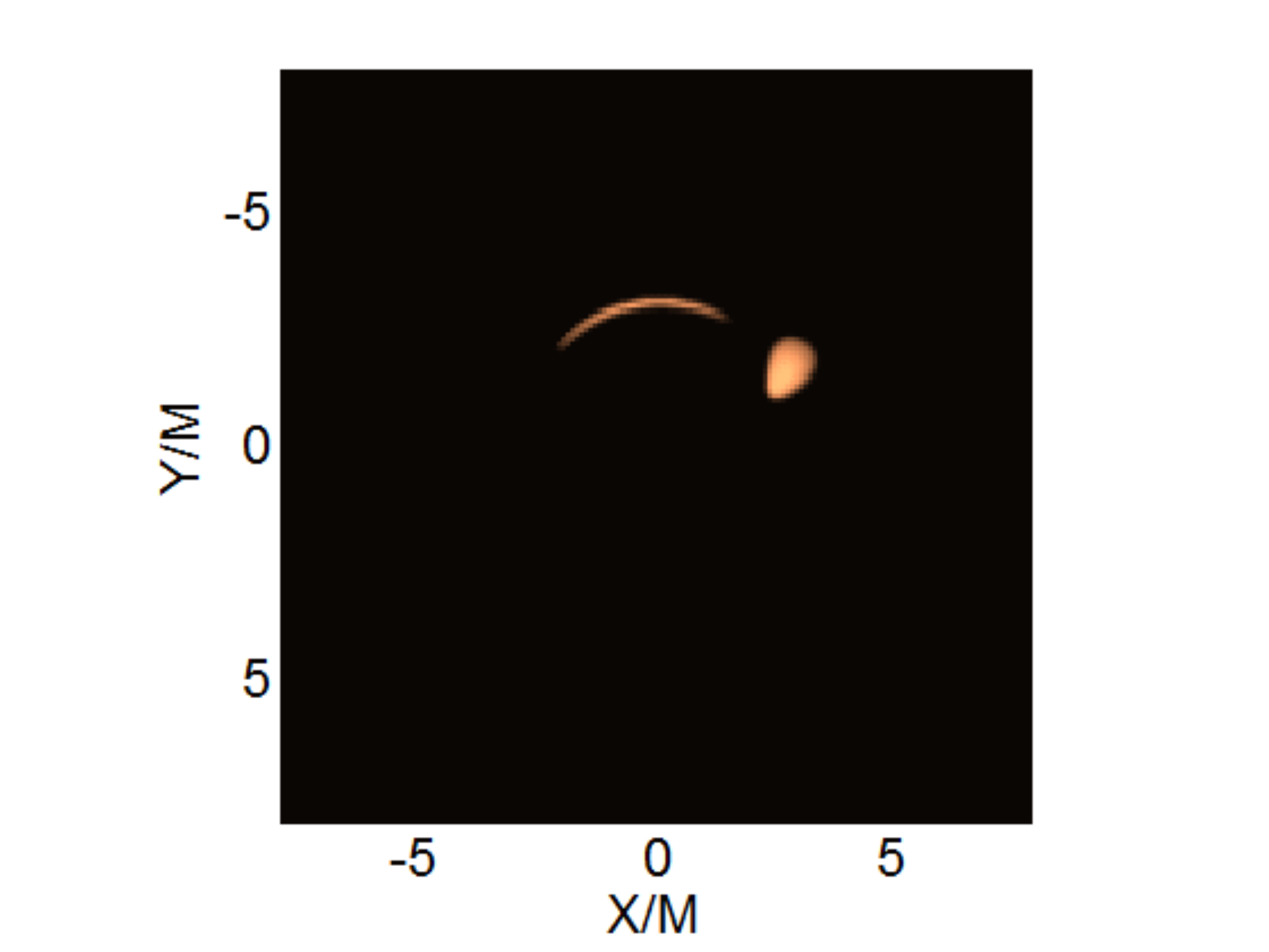} \hspace{-1cm}
\includegraphics[type=pdf,ext=.pdf,read=.pdf,width=6.5cm]{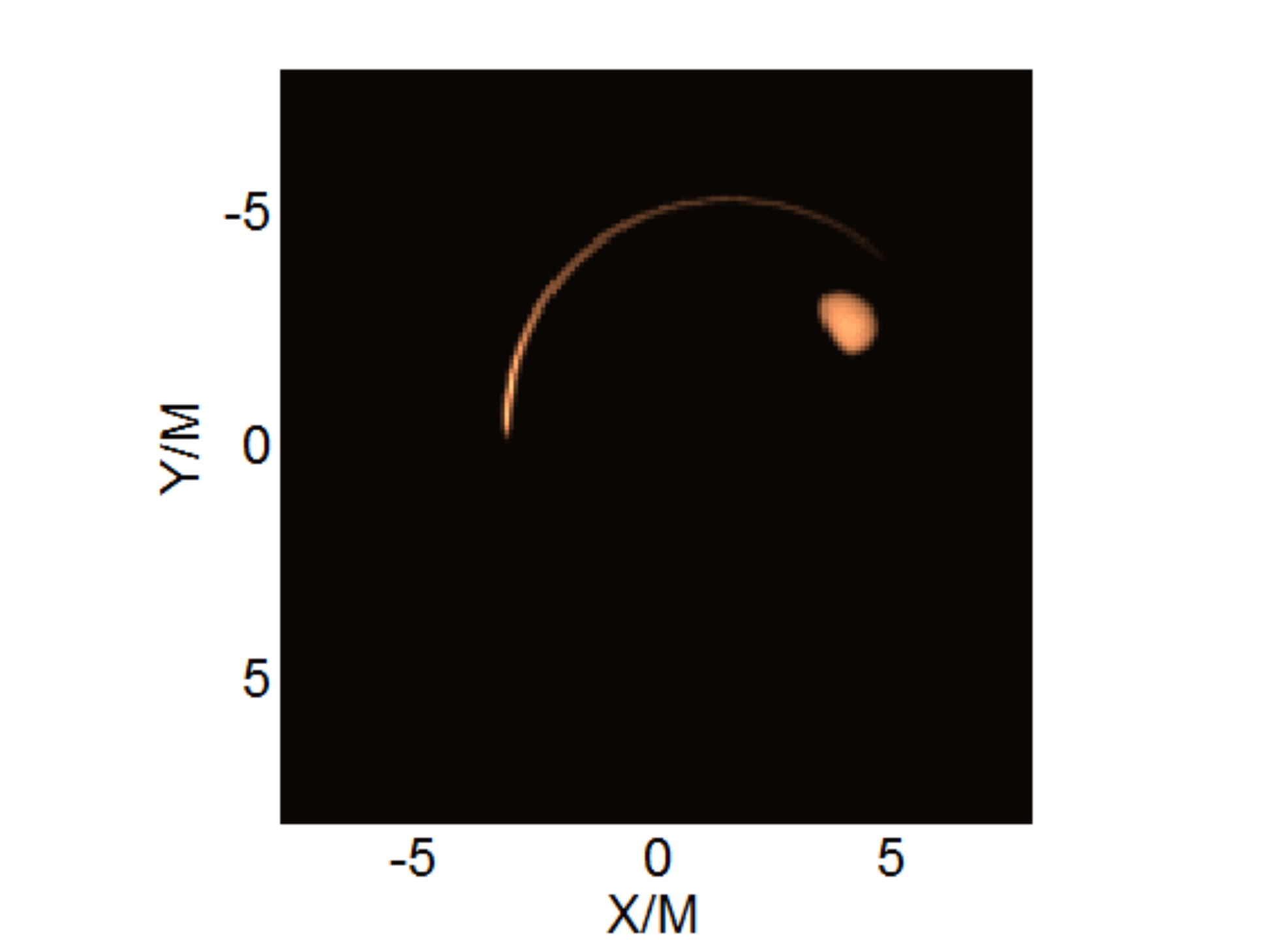} \hspace{-1cm}
\includegraphics[type=pdf,ext=.pdf,read=.pdf,width=6.5cm]{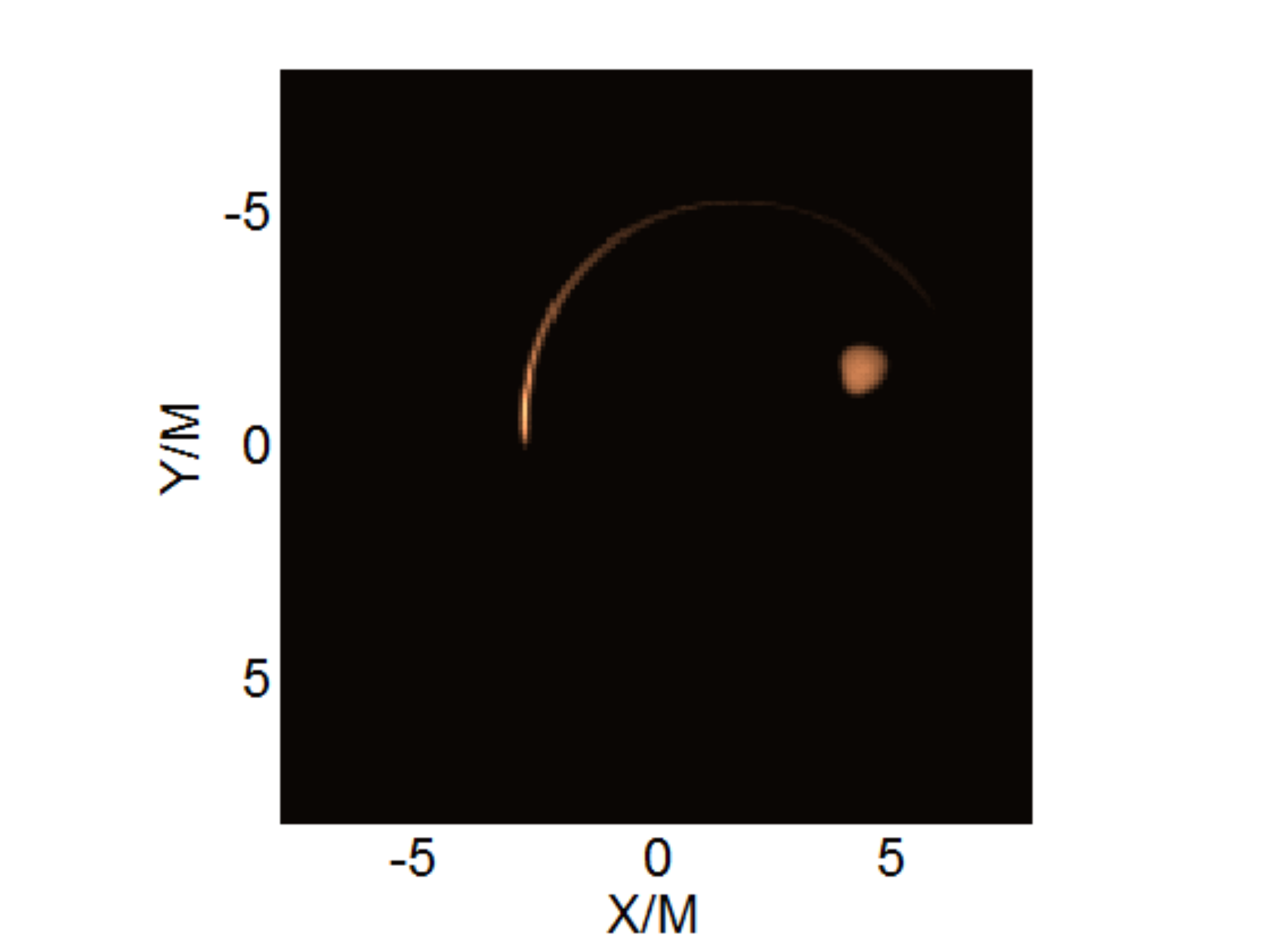} \\ \vspace{0.5cm}
\includegraphics[type=pdf,ext=.pdf,read=.pdf,width=6.5cm]{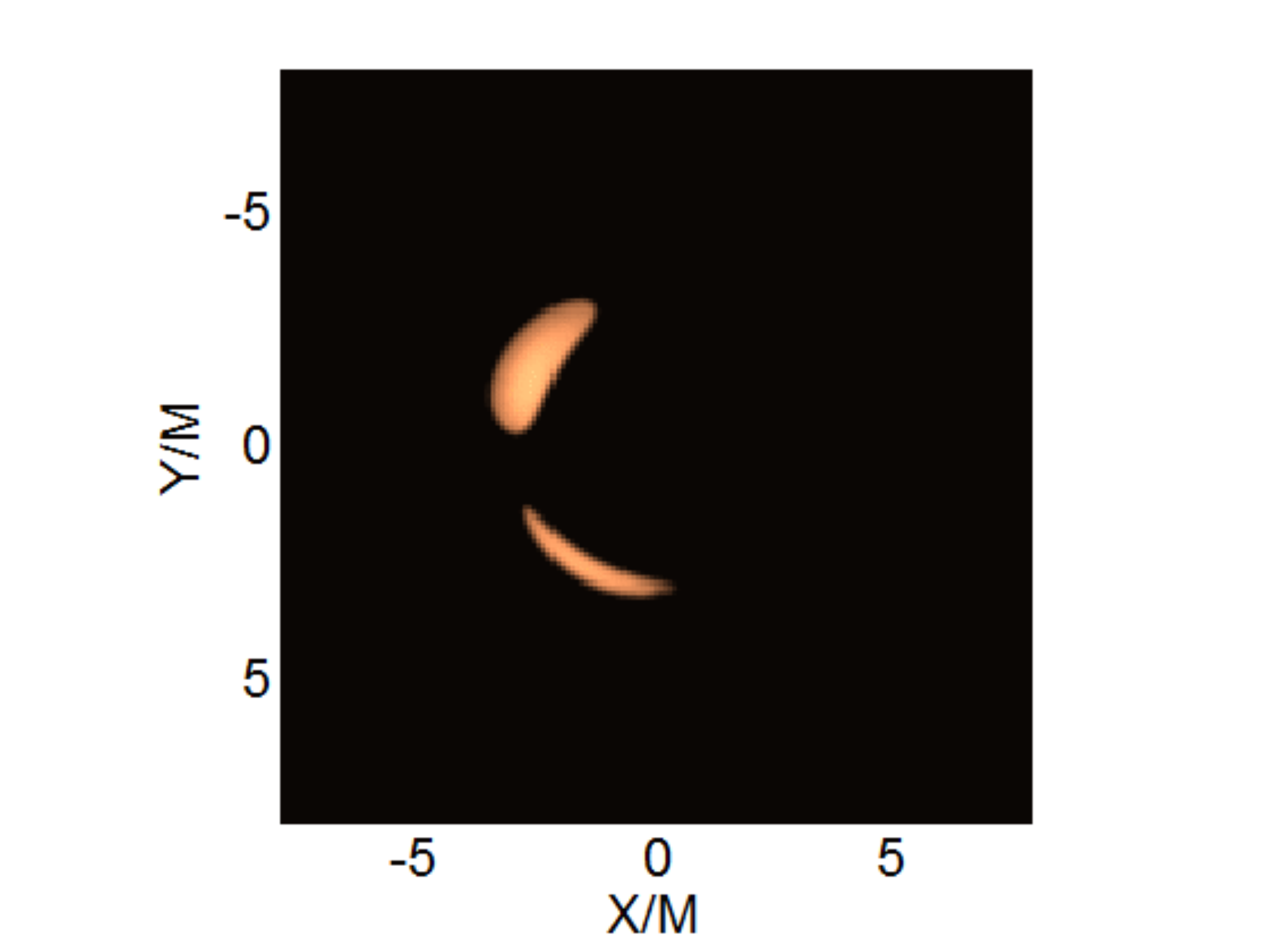} \hspace{-1cm}
\includegraphics[type=pdf,ext=.pdf,read=.pdf,width=6.5cm]{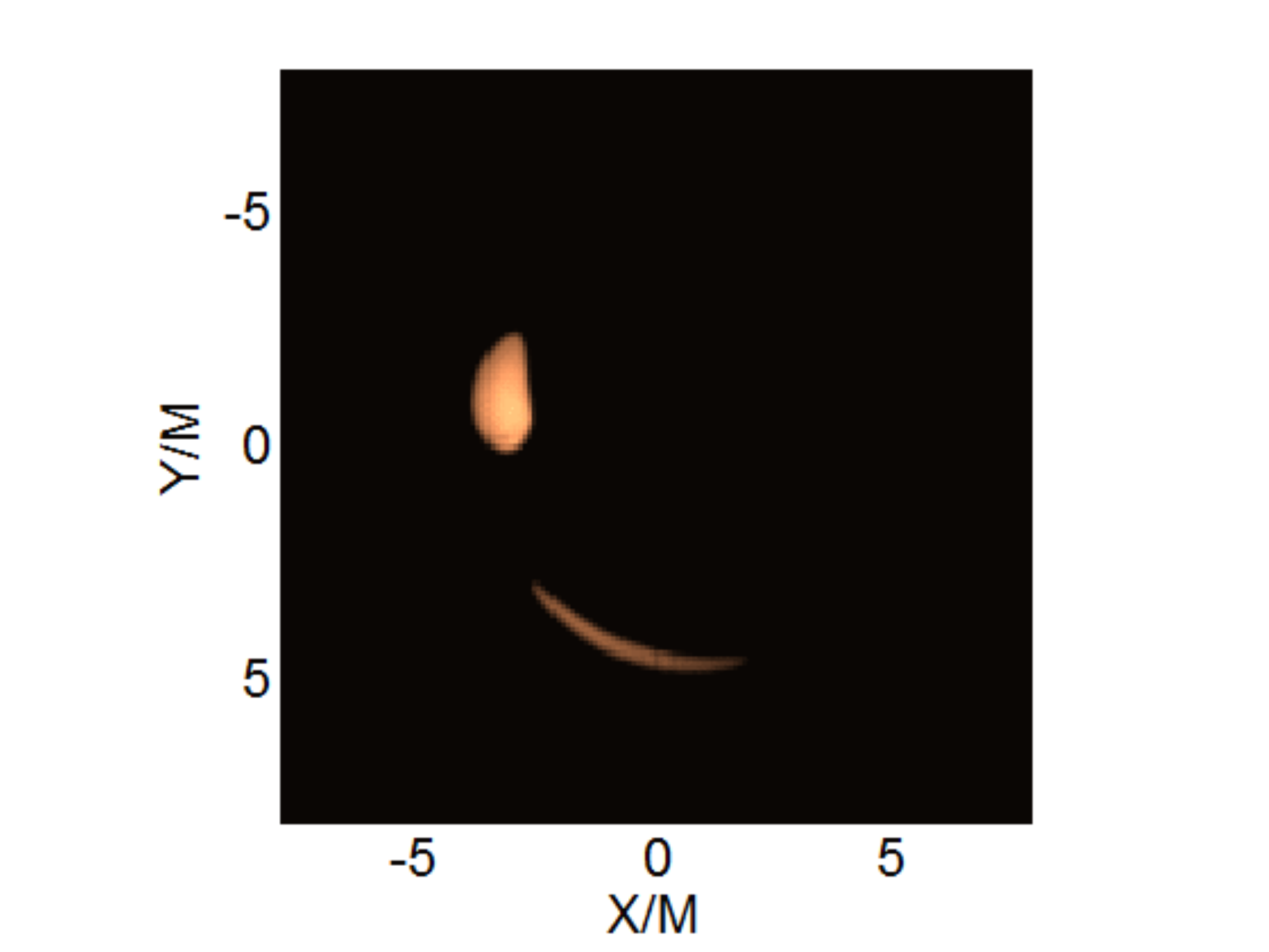} \hspace{-1cm}
\includegraphics[type=pdf,ext=.pdf,read=.pdf,width=6.5cm]{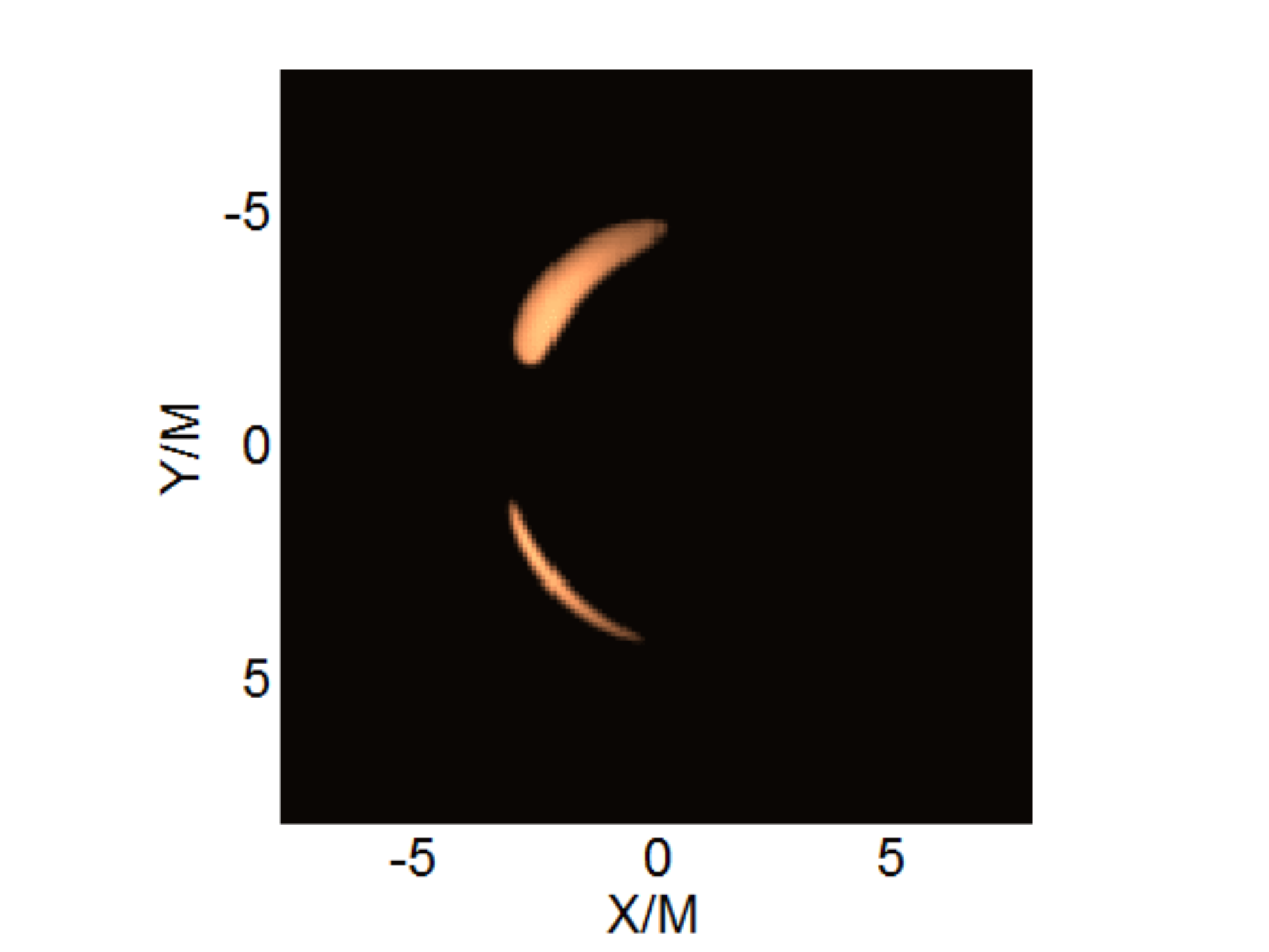} \\ \vspace{0.5cm}
\includegraphics[type=pdf,ext=.pdf,read=.pdf,width=6.5cm]{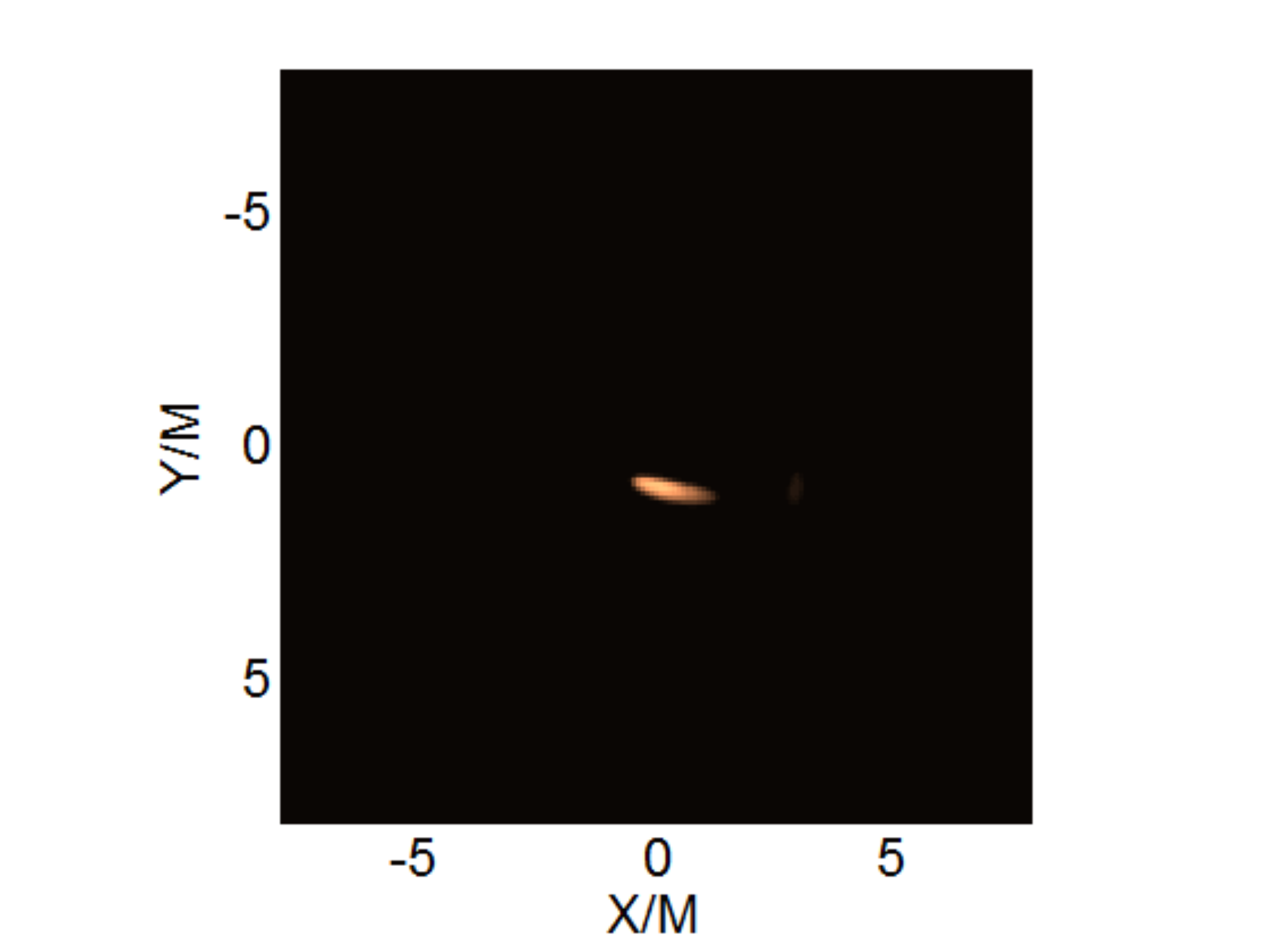} \hspace{-1cm}
\includegraphics[type=pdf,ext=.pdf,read=.pdf,width=6.5cm]{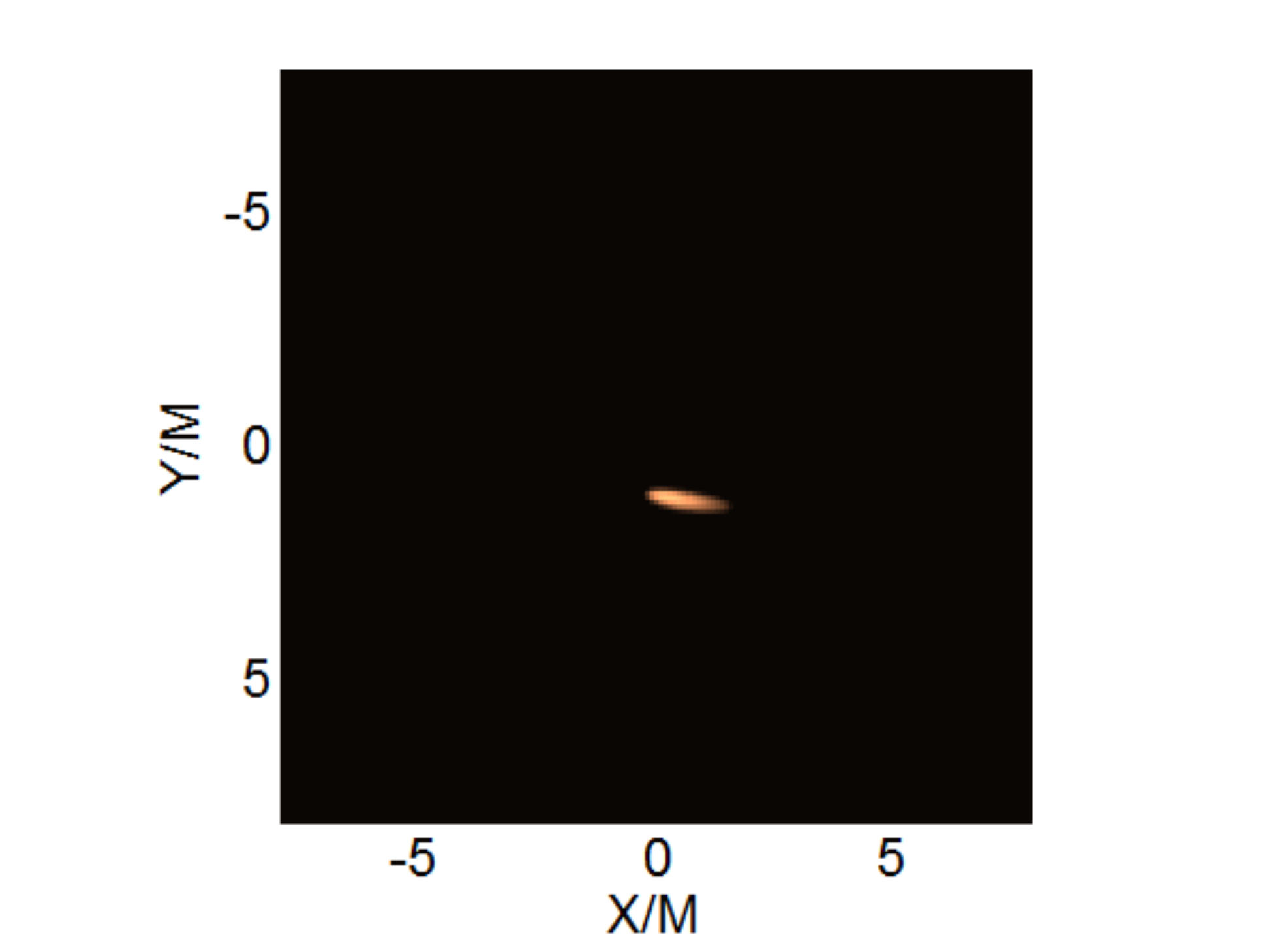} \hspace{-1cm}
\includegraphics[type=pdf,ext=.pdf,read=.pdf,width=6.5cm]{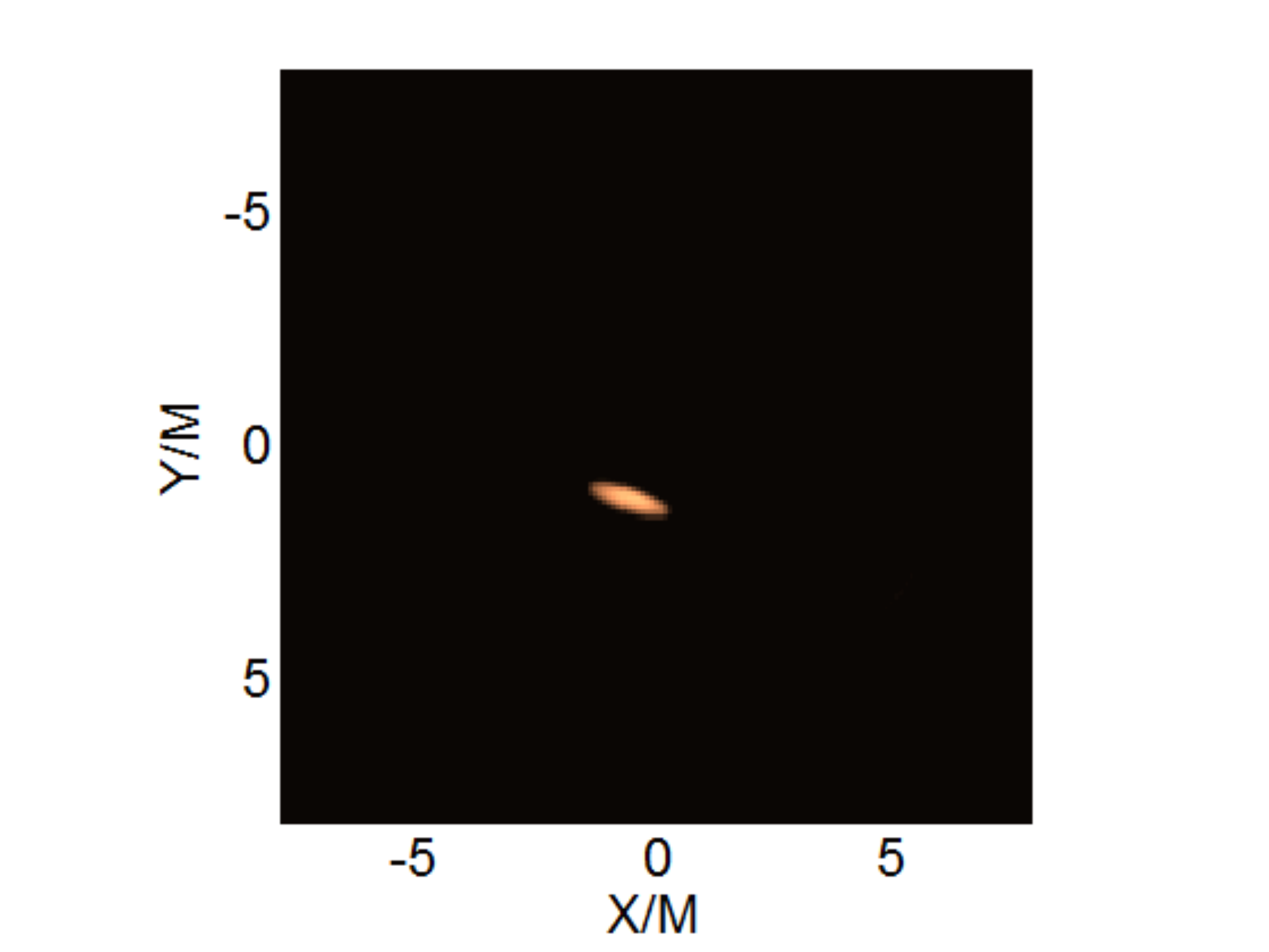} \\ \vspace{0.5cm}
\includegraphics[type=pdf,ext=.pdf,read=.pdf,width=6.5cm]{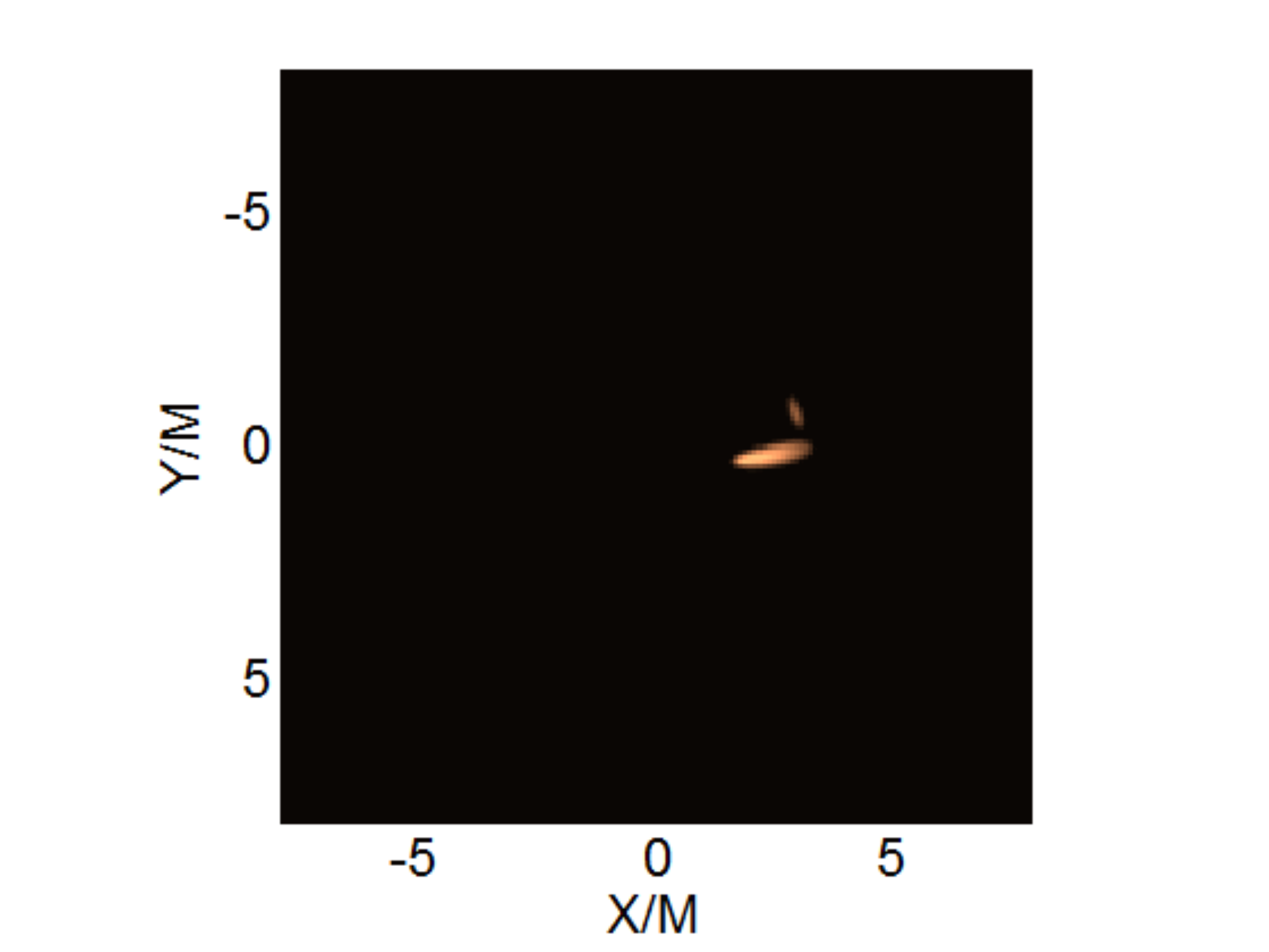} \hspace{-1cm}
\includegraphics[type=pdf,ext=.pdf,read=.pdf,width=6.5cm]{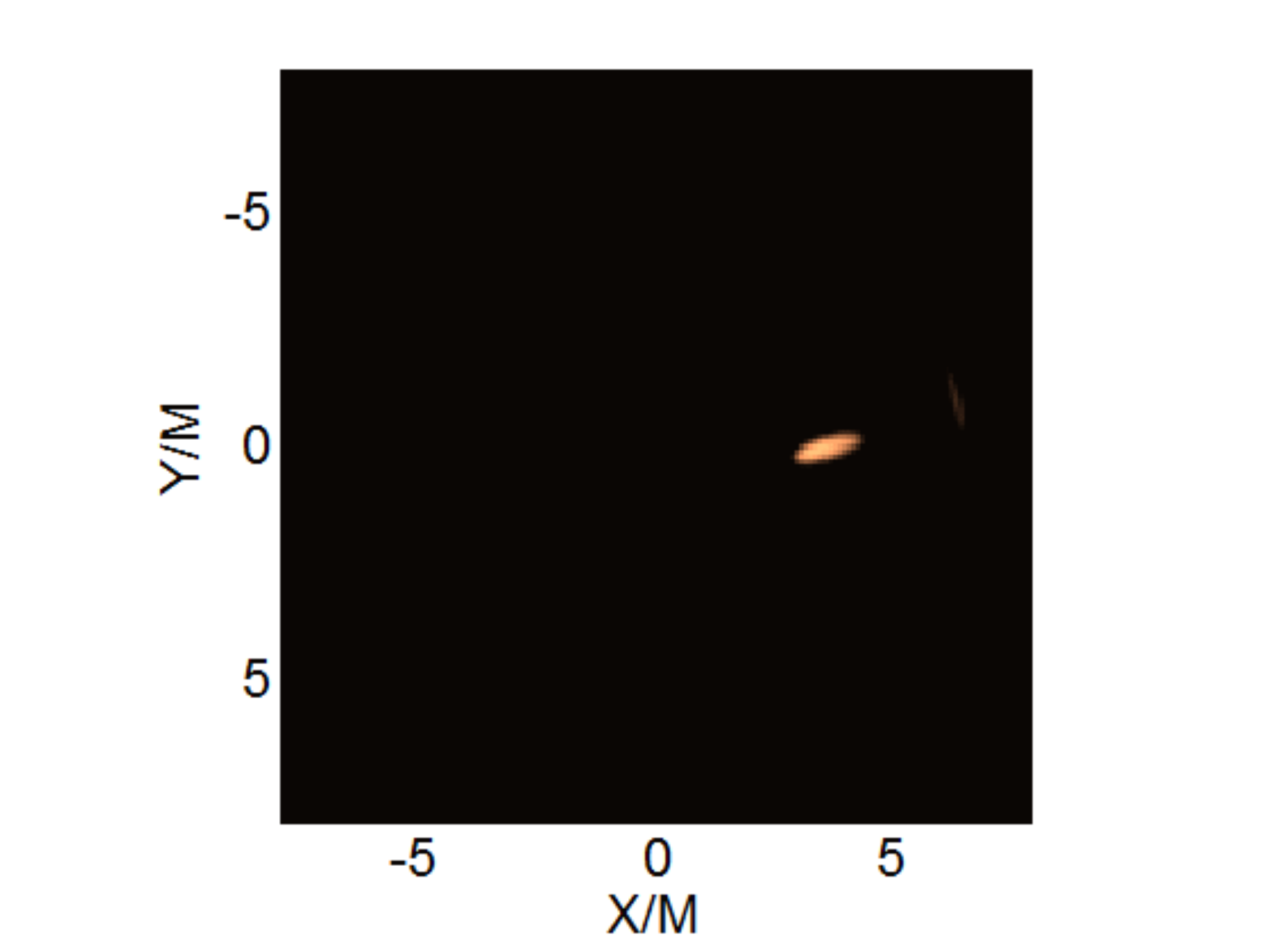} \hspace{-1cm}
\includegraphics[type=pdf,ext=.pdf,read=.pdf,width=6.5cm]{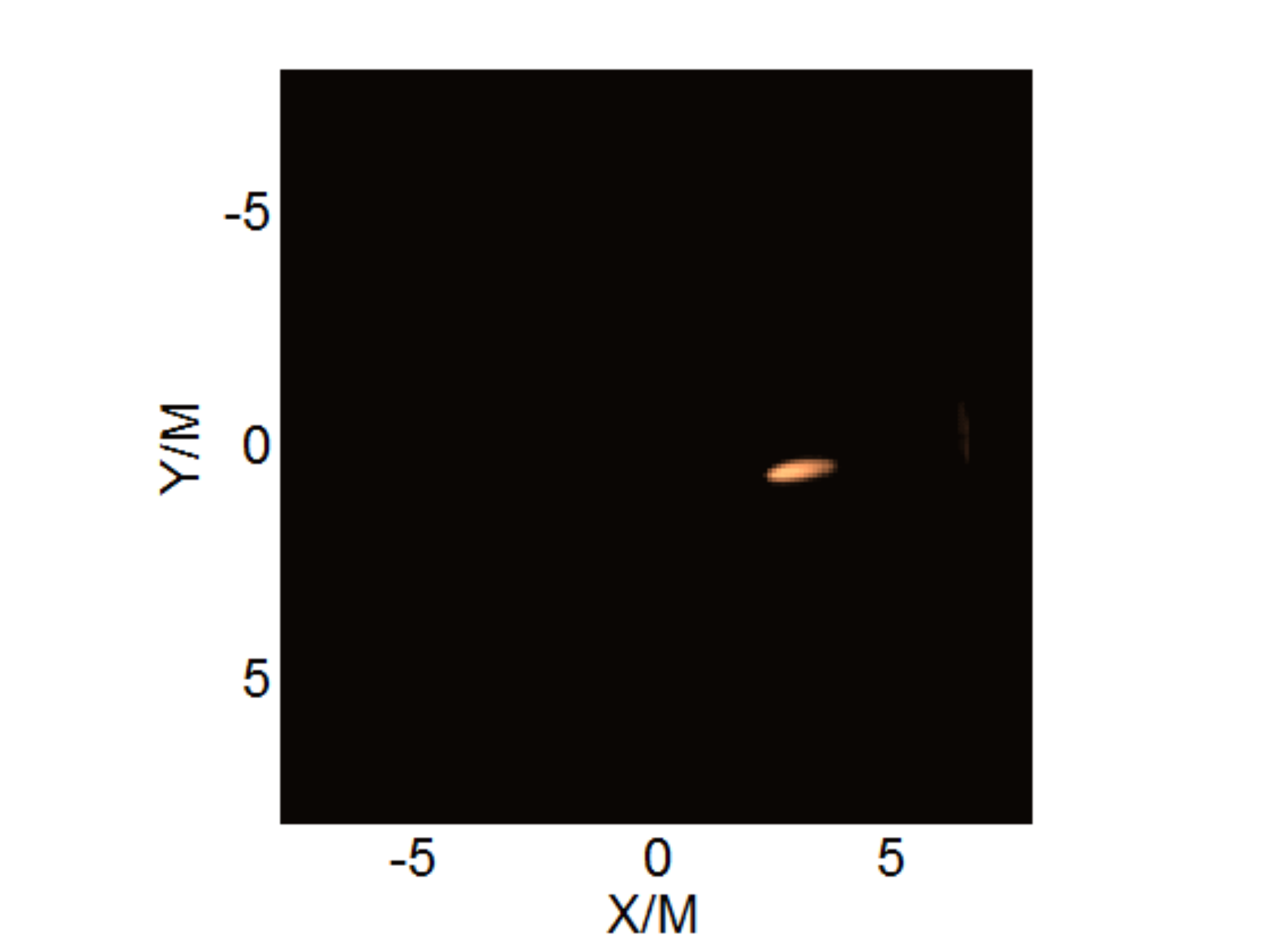} \\
\end{center}
\caption{Left panels: snapshots of a hot spot orbiting a WH at the ISCO. Central 
panels: snapshots of a hot spot orbiting a Kerr BH with spin parameter 
$a_* = 0.883911$ at the ISCO; the value of the spin parameter has been chosen 
to have the same orbital frequency as that of the hot spot orbiting the ISCO 
of the WH. Right panel: snapshots of a hot spot orbiting a Kerr BH with spin 
parameter $a_* = 0.99$ at the radius with Keplerian orbital frequency equal to 
the frequency of the hot spots of the other two cases. The time interval 
between two adjacent panels in the same column is $T/4$, where $T$ is the 
hot spot orbital period. In all these simulations, the inclination angle of the hot 
spot orbital plane with respect to the line of sight of the observer is $i = 60^\circ$ 
and the hot spot radius is $R_{\rm spot} = 0.15 \, M$. See the text for more details.}
\label{f2}
\end{figure*}

Fig.~\ref{f1} shows the light curves (total and primary image light curves,
respectively with blue-solid and red-dashed lines) of hot spots orbiting a WH 
(top panels), a Kerr BH at the ISCO radius (central panels), and a Kerr BH with
spin parameter $a_* = 0.99$ (bottom panels). The left panels correspond to hot 
spots with an angular frequency equal to that of a hot spot around a WH at 
the ISCO radius, $r_{\rm WH} = 2\, M$. The right panels correspond instead to 
hot spots with an angular frequency equal to that of a hot spot around a WH 
at the radius $r_{\rm WH} = 3\, M$. One should thus compare the light curves in 
the same column. The top panels are for the WH case with  $r_{\rm WH} = 2\, M$ 
(left panel) and $3\, M$ (right panel). The central panels 
show the light curves of a hot spot at the ISCO of a Kerr BH, whose spin parameter 
is $a_*= 0.883911$ (left panel) and 0.673917 (right panel). In the bottom panels,
there are the hot spot light curves around a Kerr BH with $a_* = 0.99$: the hot 
spot orbital radius is respectively $r_{\rm BH} = 2.3807\, M$ (left panel) and 
$3.3973\, M$ (right panel).

The most important 
difference between the WH and BH cases is that in the BH light curves there is
a small bump marking the maximum intensity of the secondary image light
curve, which is instead absent in the WH light curve. The effect is more 
pronounced when the hot spot is closer to the compact object and tends to
disappear as the hot spot radius/frequency increases. While such a feature
in the hot spot light curve could potentially represent an observational 
signature to distinguish WHs and BHs, the actual properties of the bump due 
to the secondary image depend on the hot spot model (hot spot size, emissivity
function, etc.). In Fig.~\ref{f1}, we have considered a single hot spot disk with 
radius $R_{\rm spot} = 0.15 \, M$ and isotropic and monochromatic emission.
The strong gravitational force near the compact object typically tends to 
destroy the hot spot, which is smeared along its orbit with the result to make
the peaks of the images less well defined. Moreover, the substantial background
may hide the small bump due to the secondary image in the BH light curves.
While a final answer would require a more detailed discussion 
based on a more realistic model, from these results it seems unlikely that the 
observation of the light curve of a hot spot can be used to distinguish BHs 
and WHs, even considering that real data are usually noisy and incomplete.

\begin{figure*}
\begin{center}
\includegraphics[type=pdf,ext=.pdf,read=.pdf,width=6.5cm]{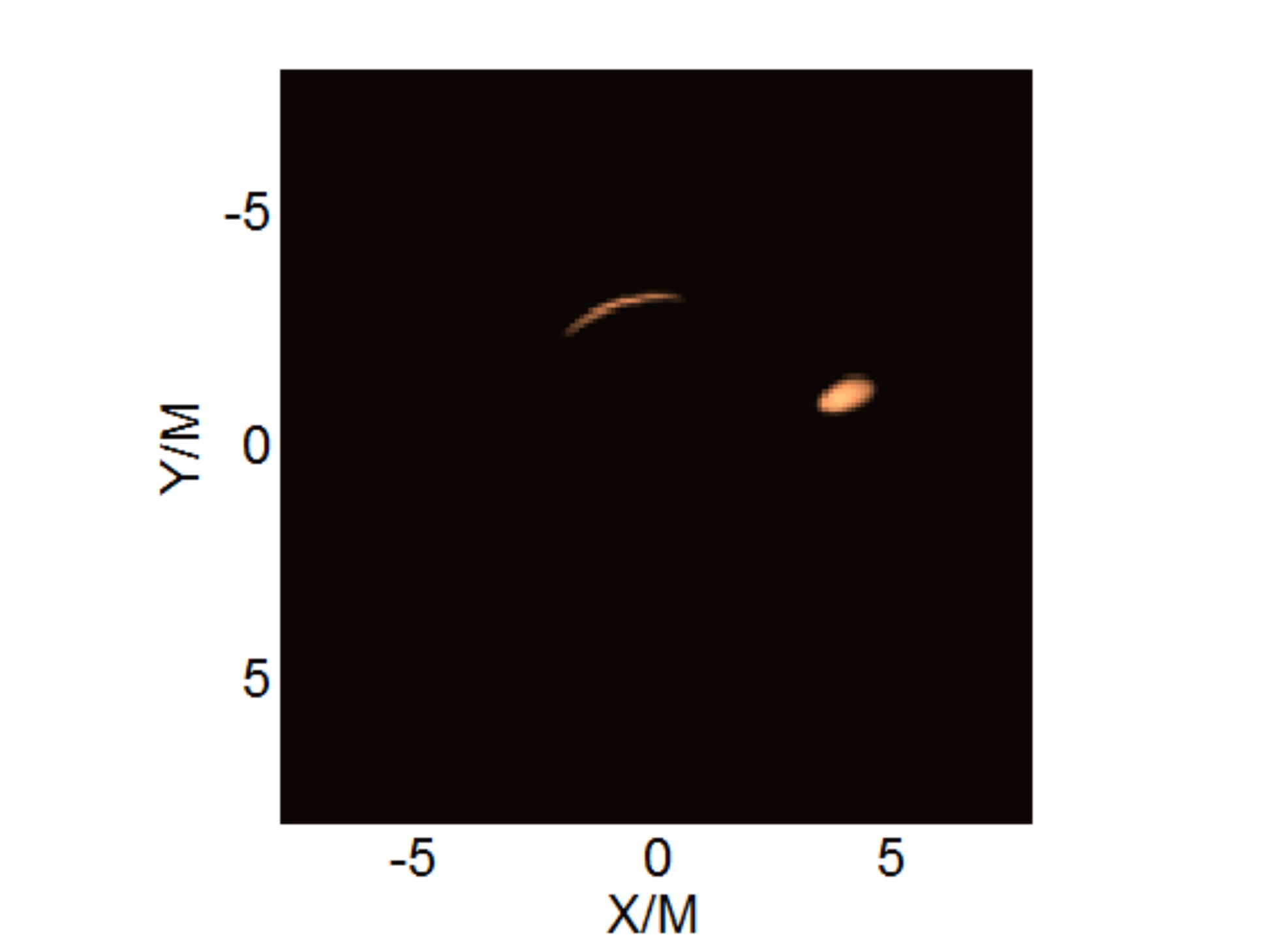} \hspace{-1cm}
\includegraphics[type=pdf,ext=.pdf,read=.pdf,width=6.5cm]{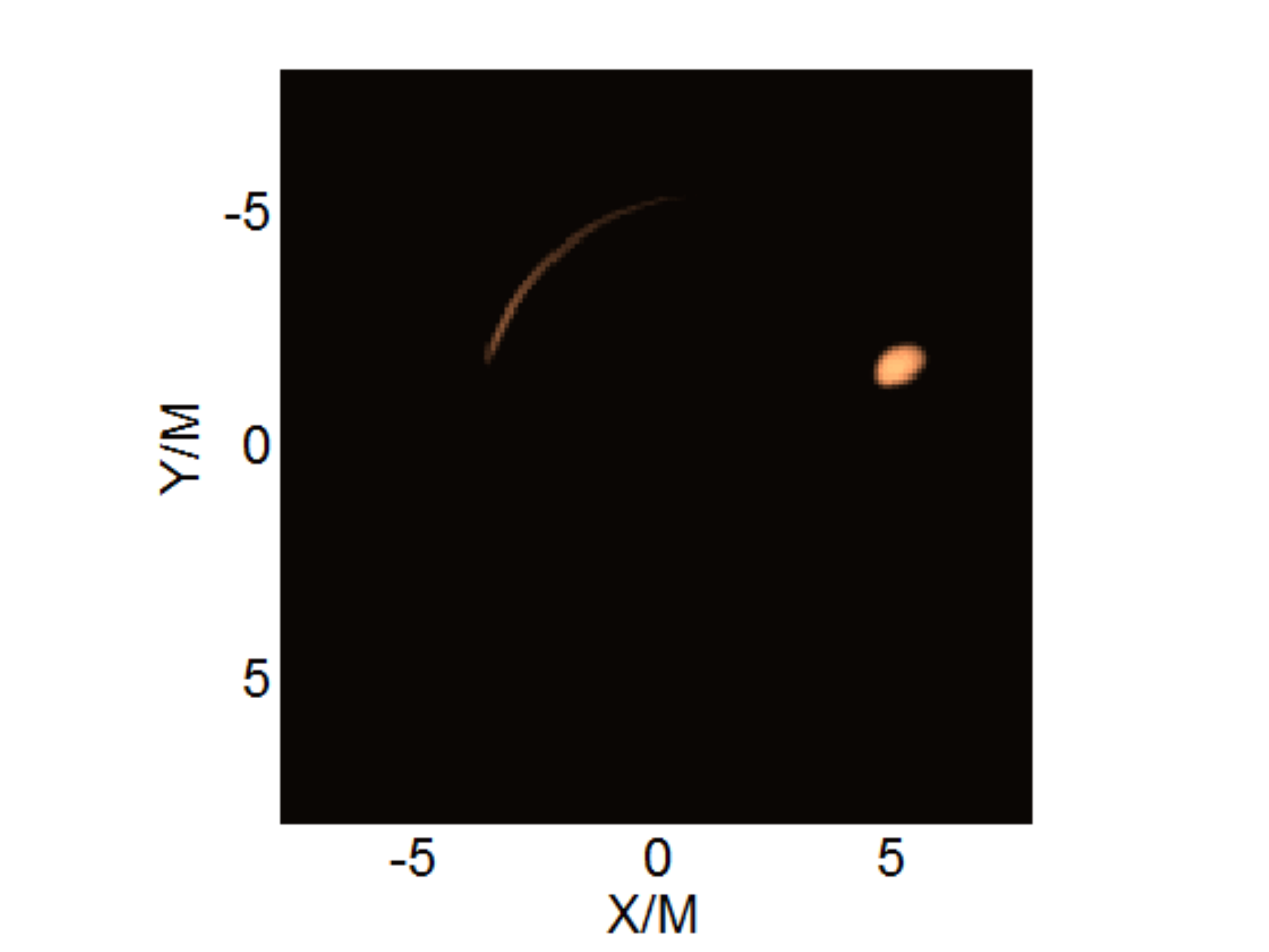} \hspace{-1cm}
\includegraphics[type=pdf,ext=.pdf,read=.pdf,width=6.5cm]{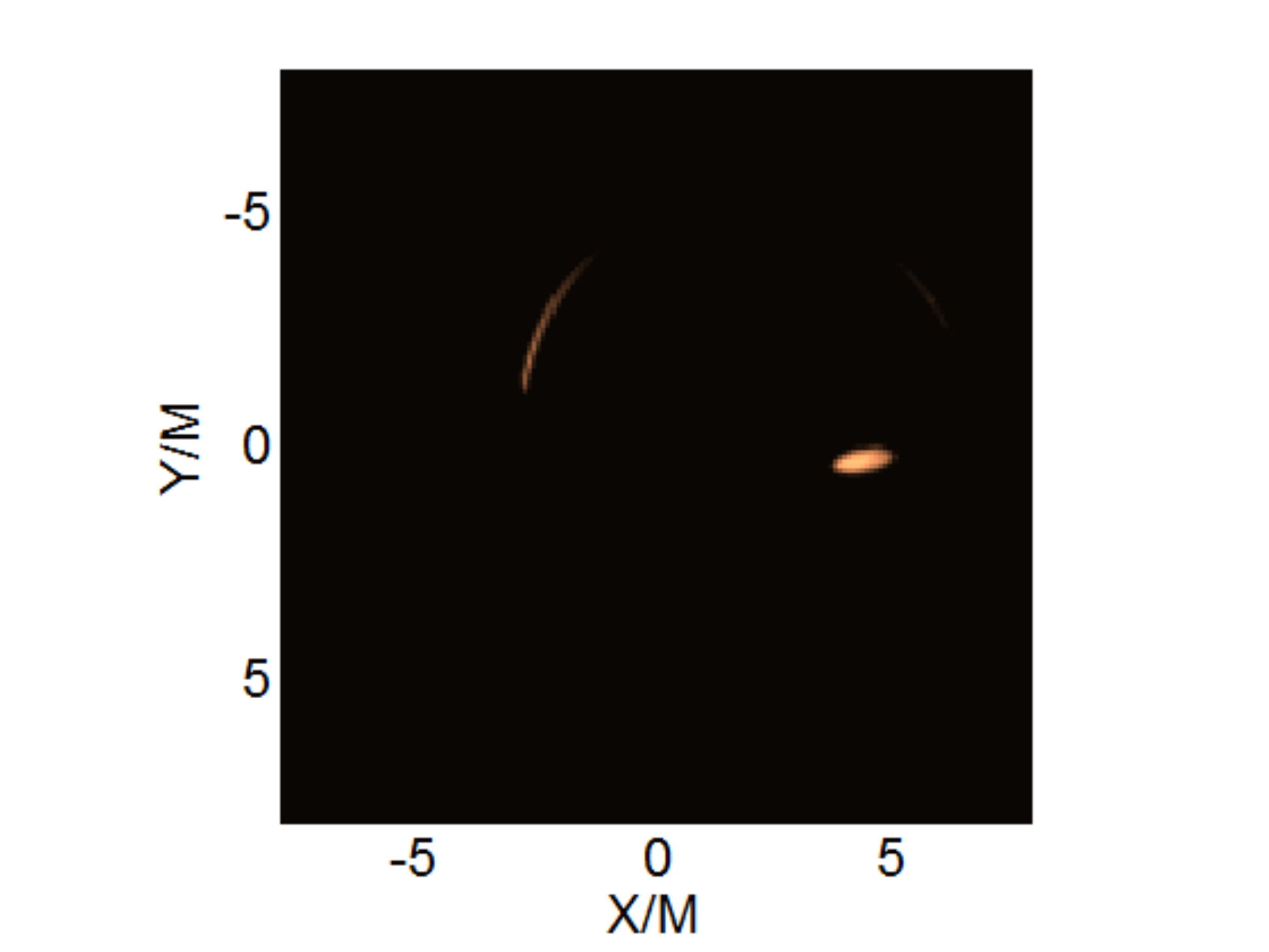} \\ \vspace{0.5cm}
\includegraphics[type=pdf,ext=.pdf,read=.pdf,width=6.5cm]{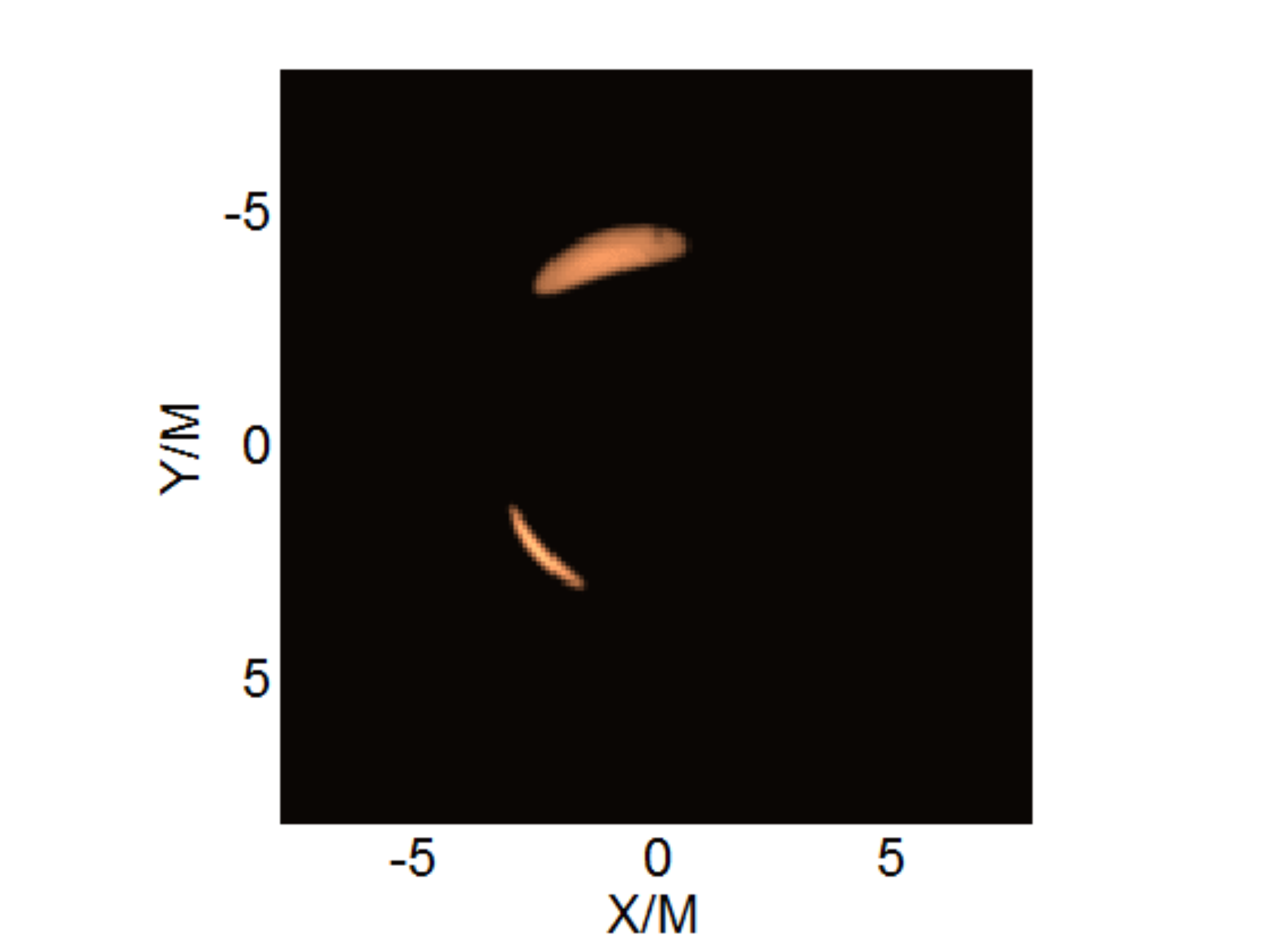} \hspace{-1cm}
\includegraphics[type=pdf,ext=.pdf,read=.pdf,width=6.5cm]{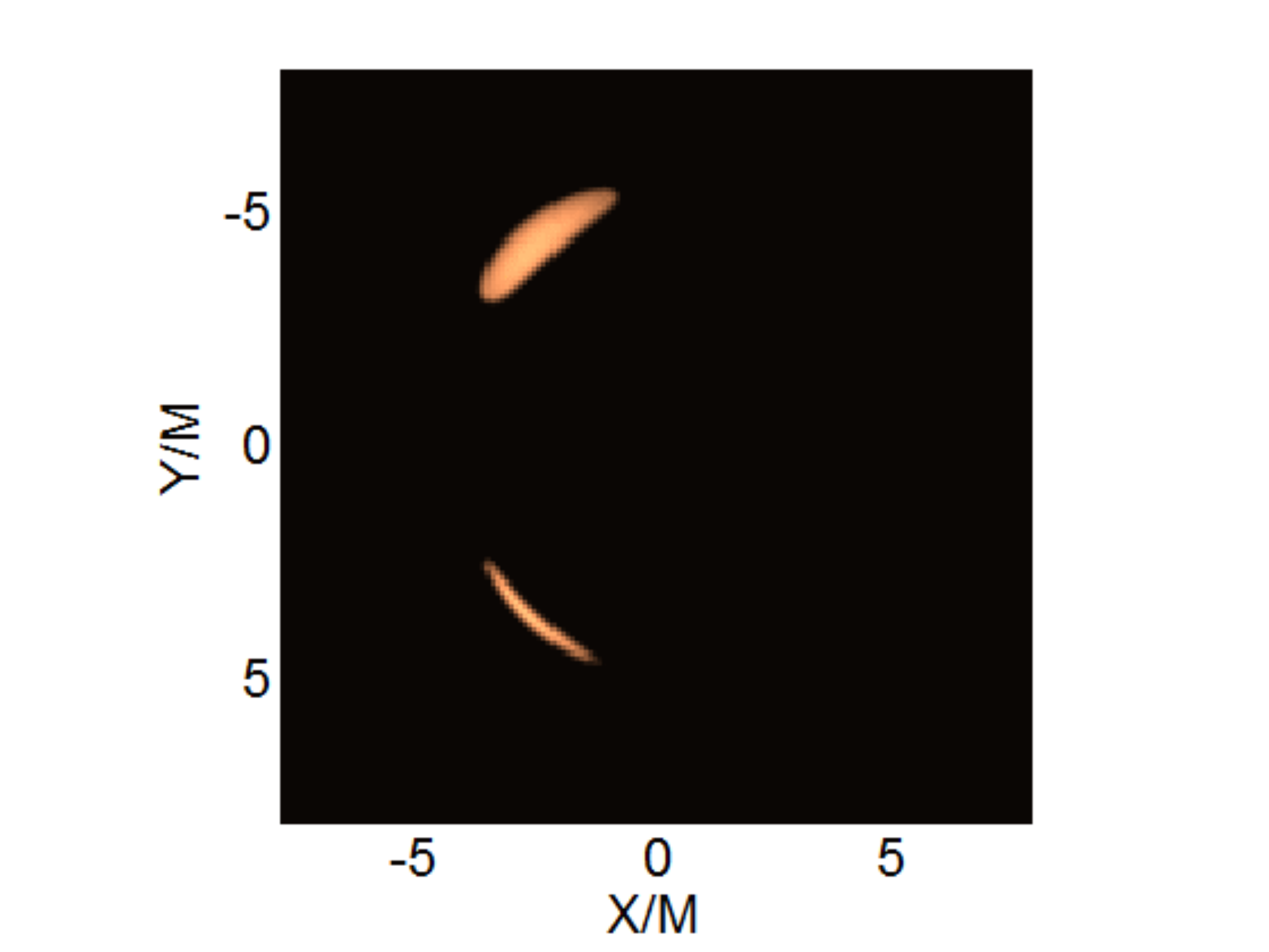} \hspace{-1cm}
\includegraphics[type=pdf,ext=.pdf,read=.pdf,width=6.5cm]{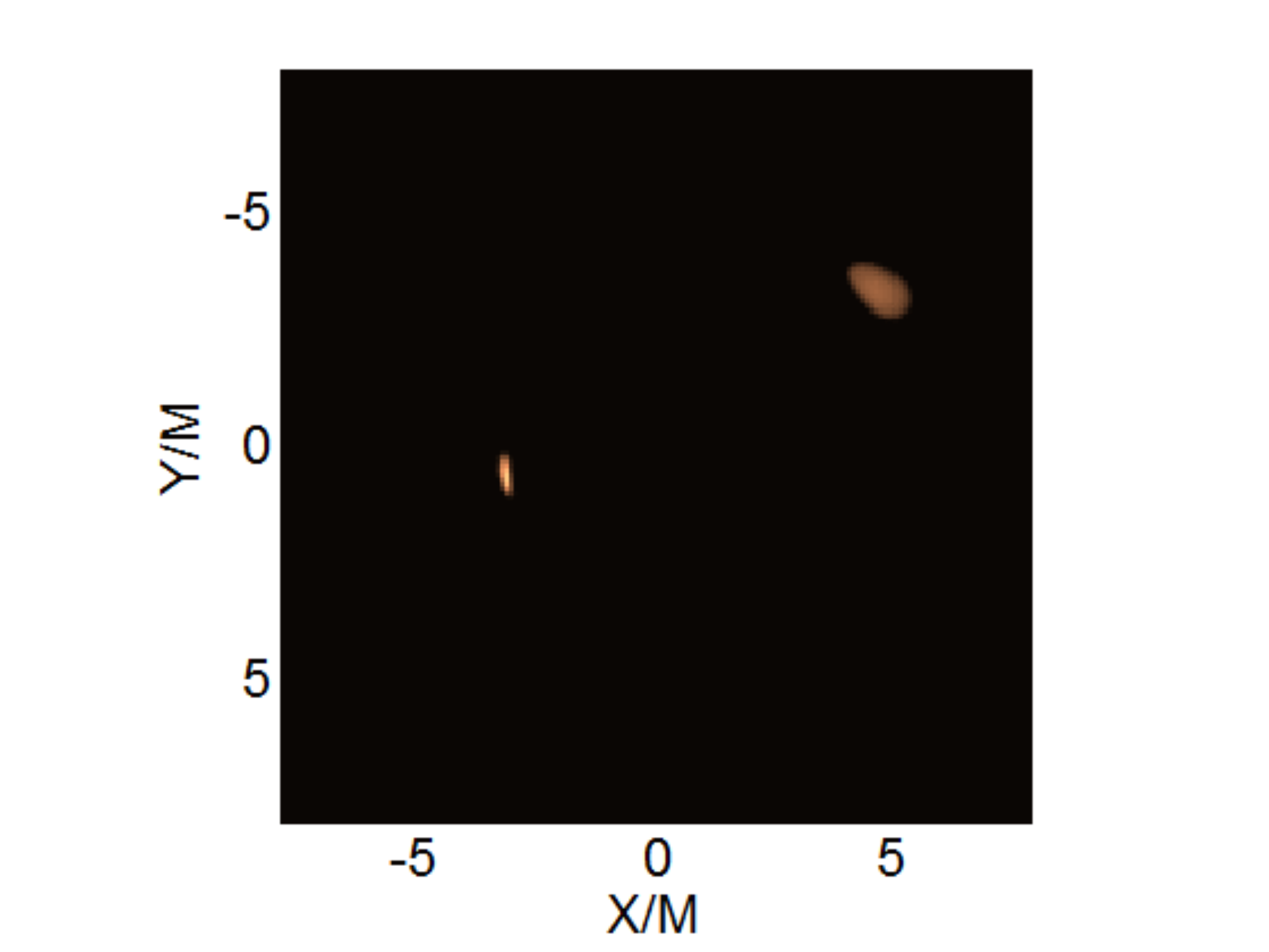} \\ \vspace{0.5cm}
\includegraphics[type=pdf,ext=.pdf,read=.pdf,width=6.5cm]{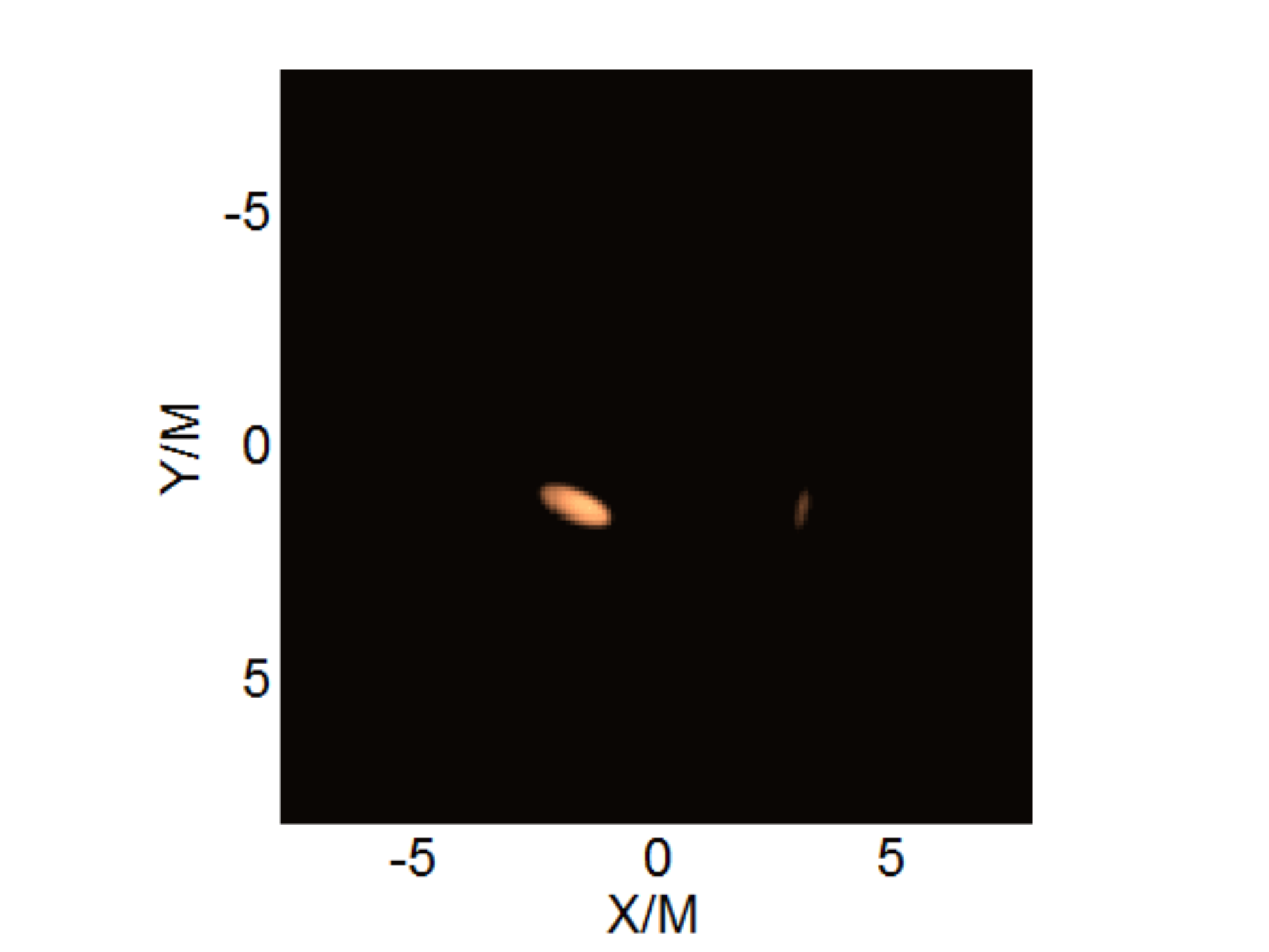} \hspace{-1cm}
\includegraphics[type=pdf,ext=.pdf,read=.pdf,width=6.5cm]{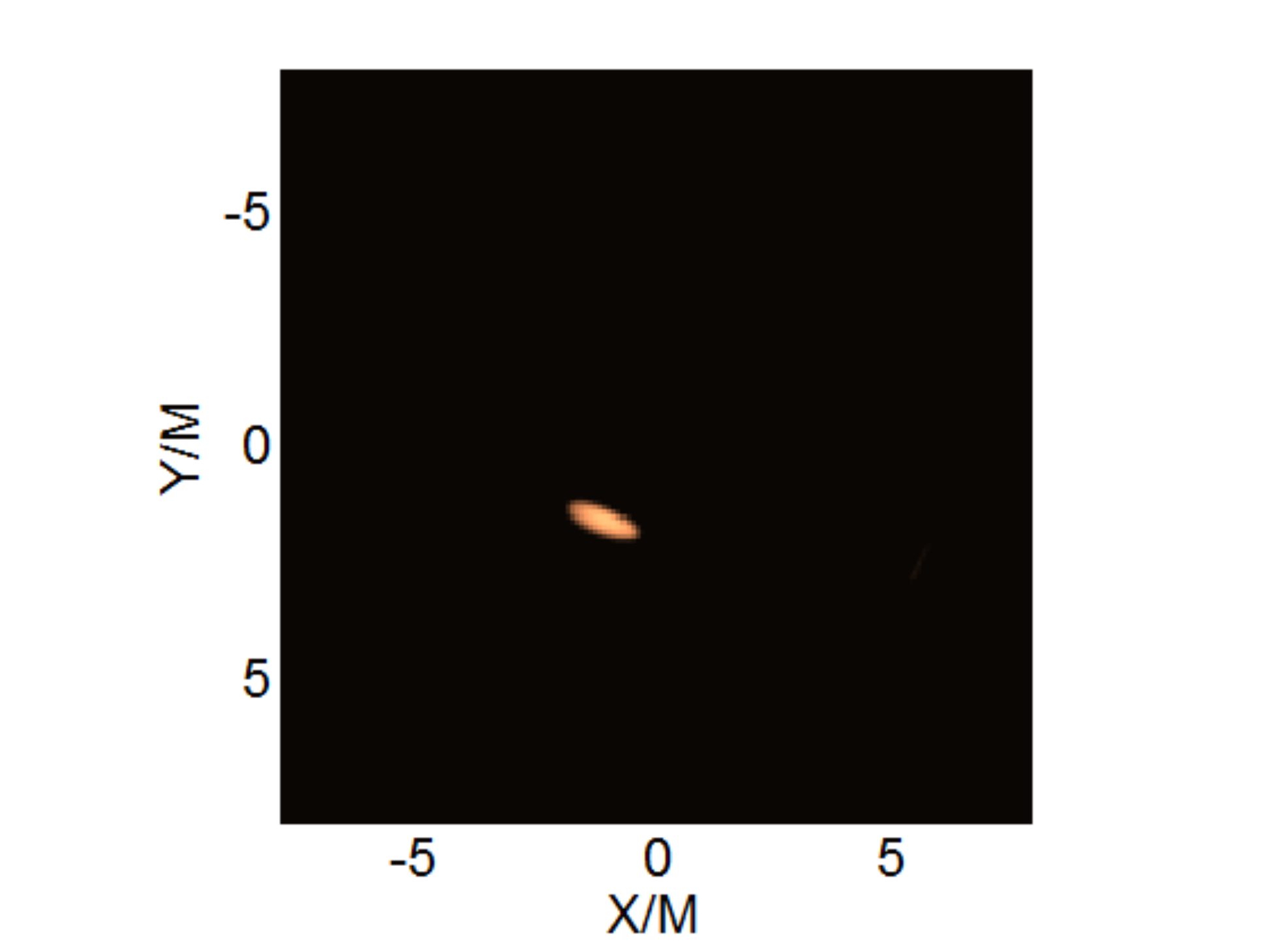} \hspace{-1cm}
\includegraphics[type=pdf,ext=.pdf,read=.pdf,width=6.5cm]{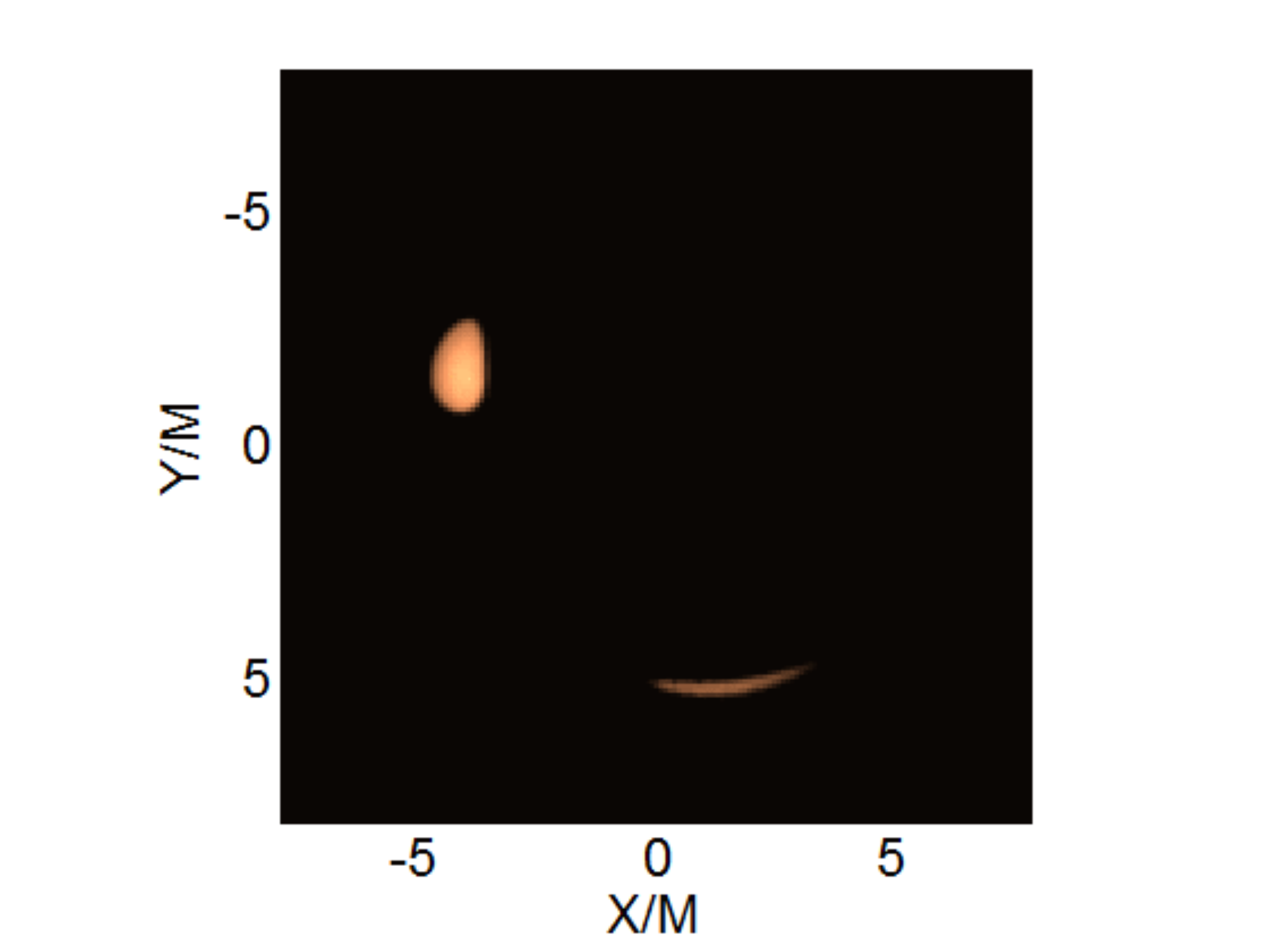} \\ \vspace{0.5cm}
\includegraphics[type=pdf,ext=.pdf,read=.pdf,width=6.5cm]{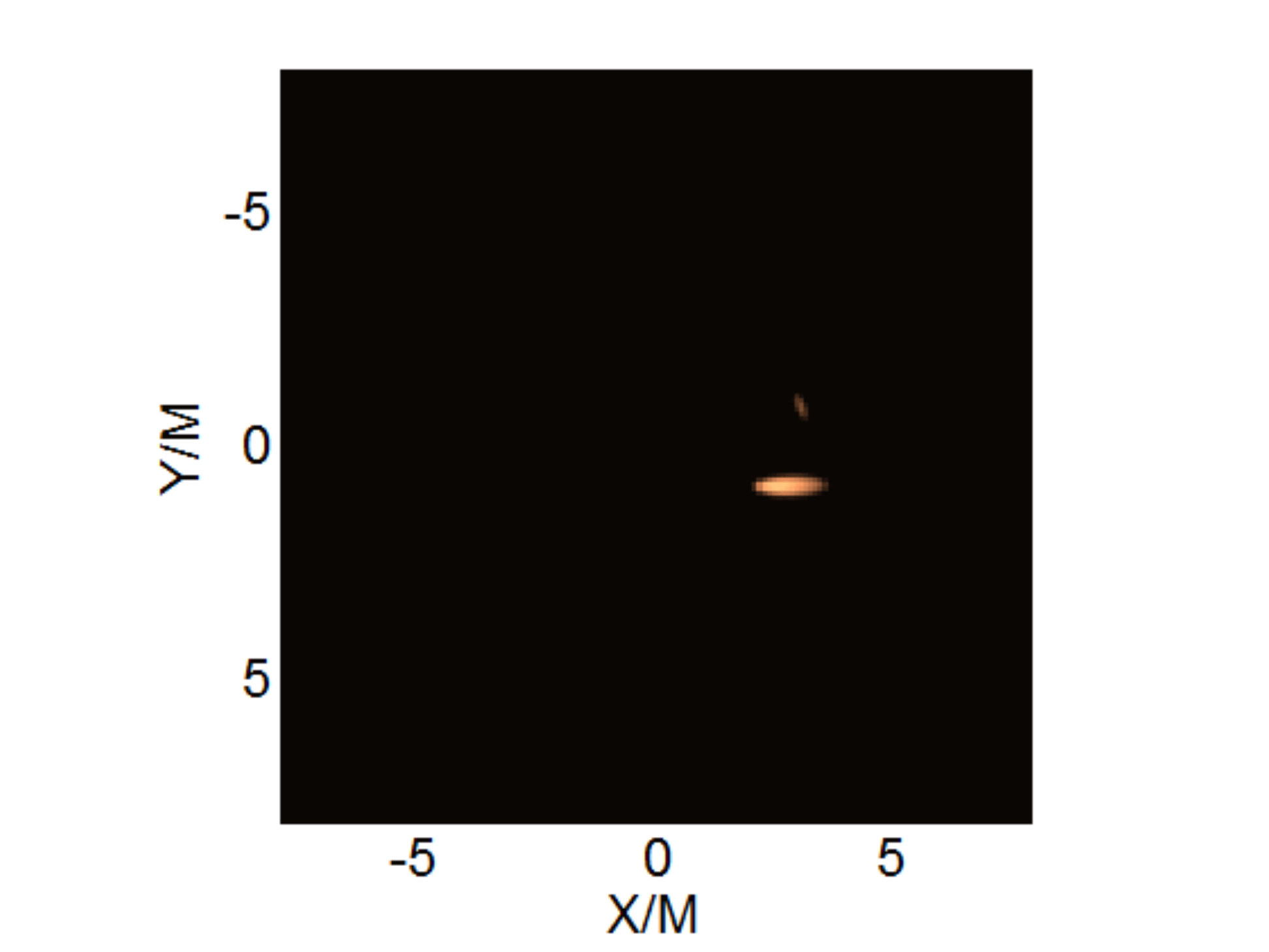} \hspace{-1cm}
\includegraphics[type=pdf,ext=.pdf,read=.pdf,width=6.5cm]{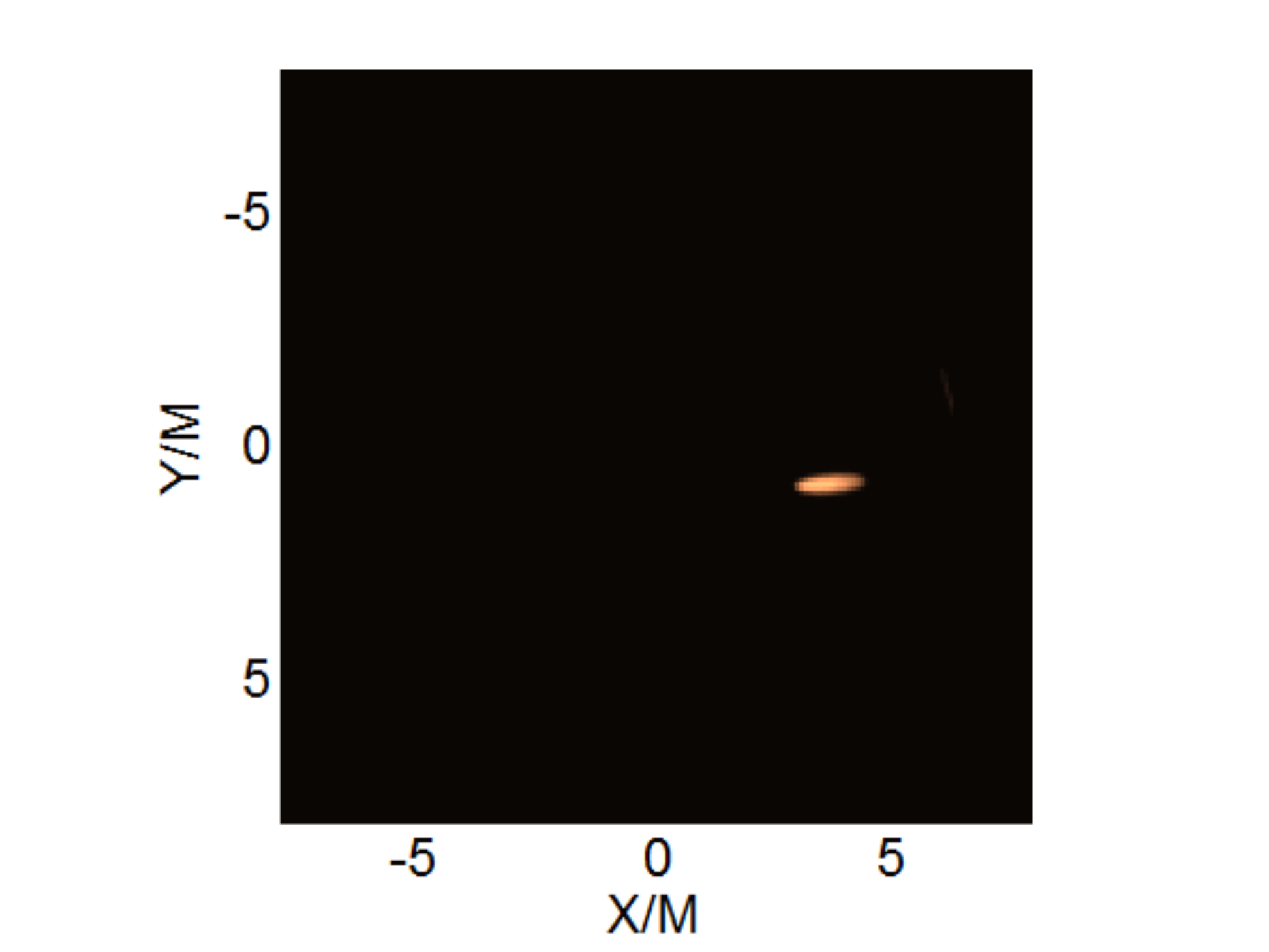} \hspace{-1cm}
\includegraphics[type=pdf,ext=.pdf,read=.pdf,width=6.5cm]{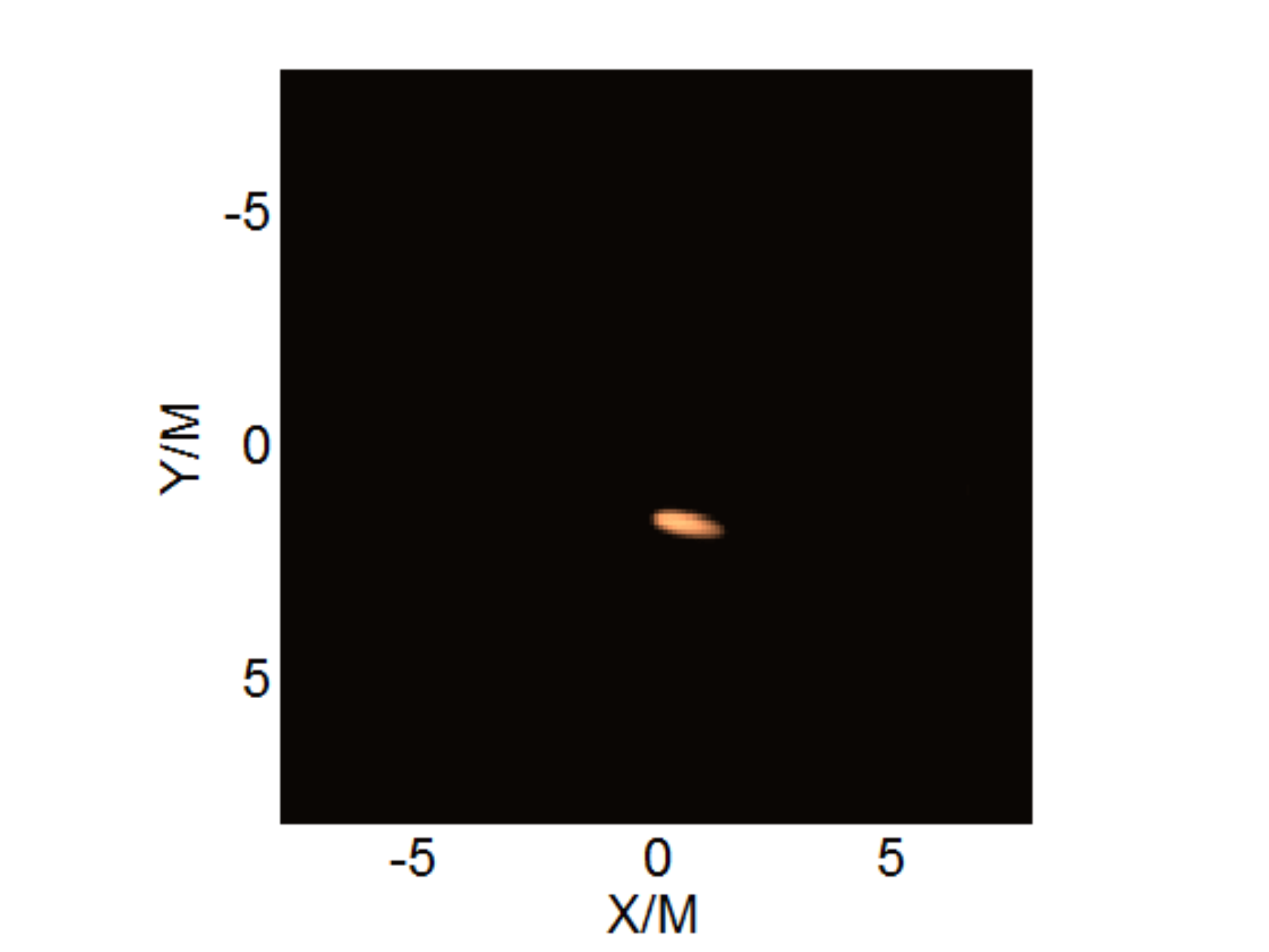} \\
\end{center}
\caption{As in Fig.~\ref{f2} for a hot spot orbiting a WH at the radius $r_{\rm WH} = 3 \, M$ 
(left panels), a Kerr BH with spin parameter $a_* = 0.673917$ at the ISCO (central 
panels; the value of the spin parameter has been chosen to have the same hot 
spot orbital period as the one around the WH), and a Kerr BH with spin parameter 
$a_*=0.99$ and at the radius with Keplerian orbital period equal to that of 
the other two cases (right panels). See the text for more details.}
\label{f3}
\end{figure*}

\begin{figure*}
\begin{center}
\vspace{0.5cm}
\includegraphics[type=pdf,ext=.pdf,read=.pdf,width=7.5cm]{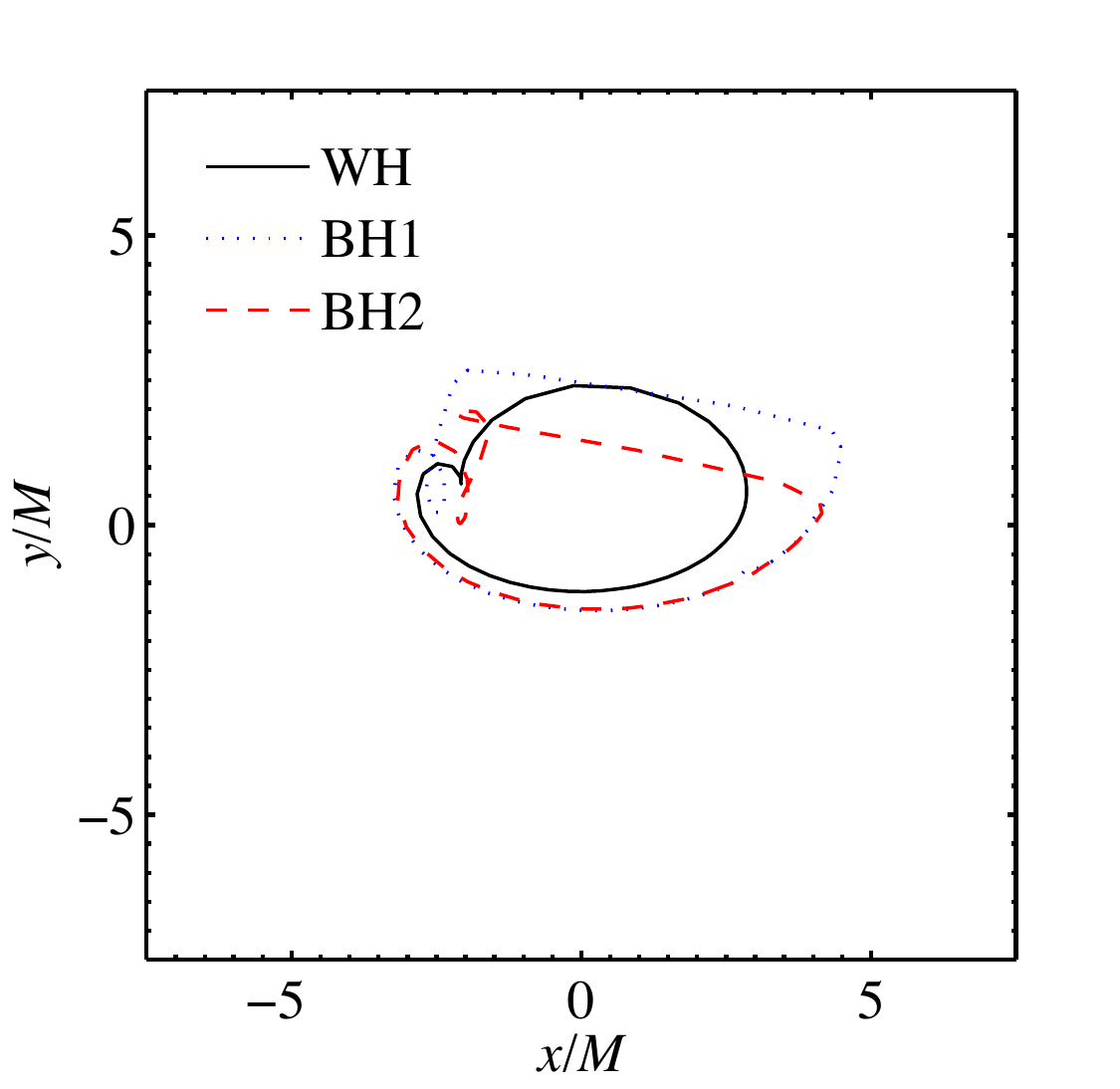} \hspace{0.5cm}
\includegraphics[type=pdf,ext=.pdf,read=.pdf,width=7.5cm]{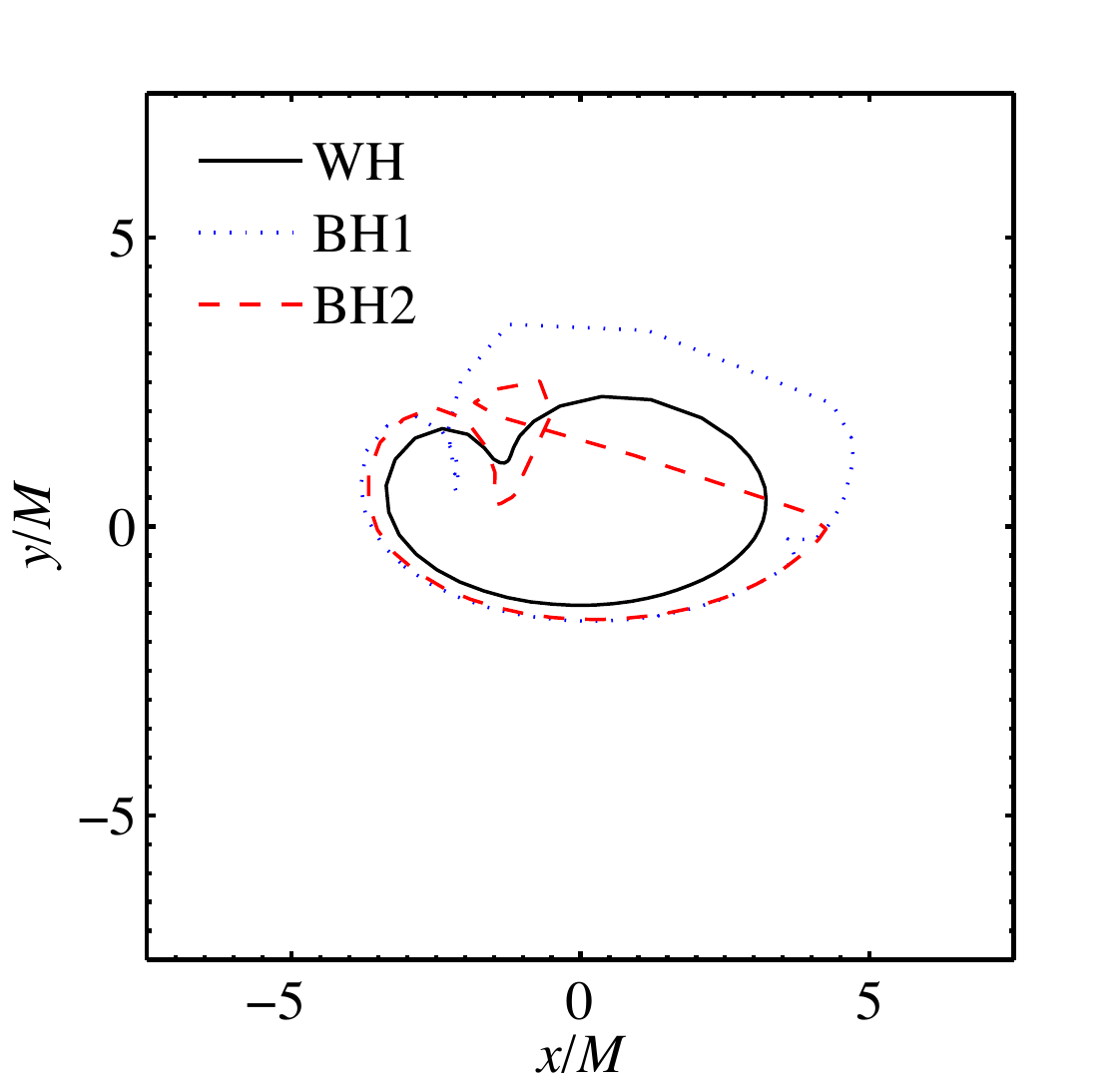} \\
\includegraphics[type=pdf,ext=.pdf,read=.pdf,width=7.5cm]{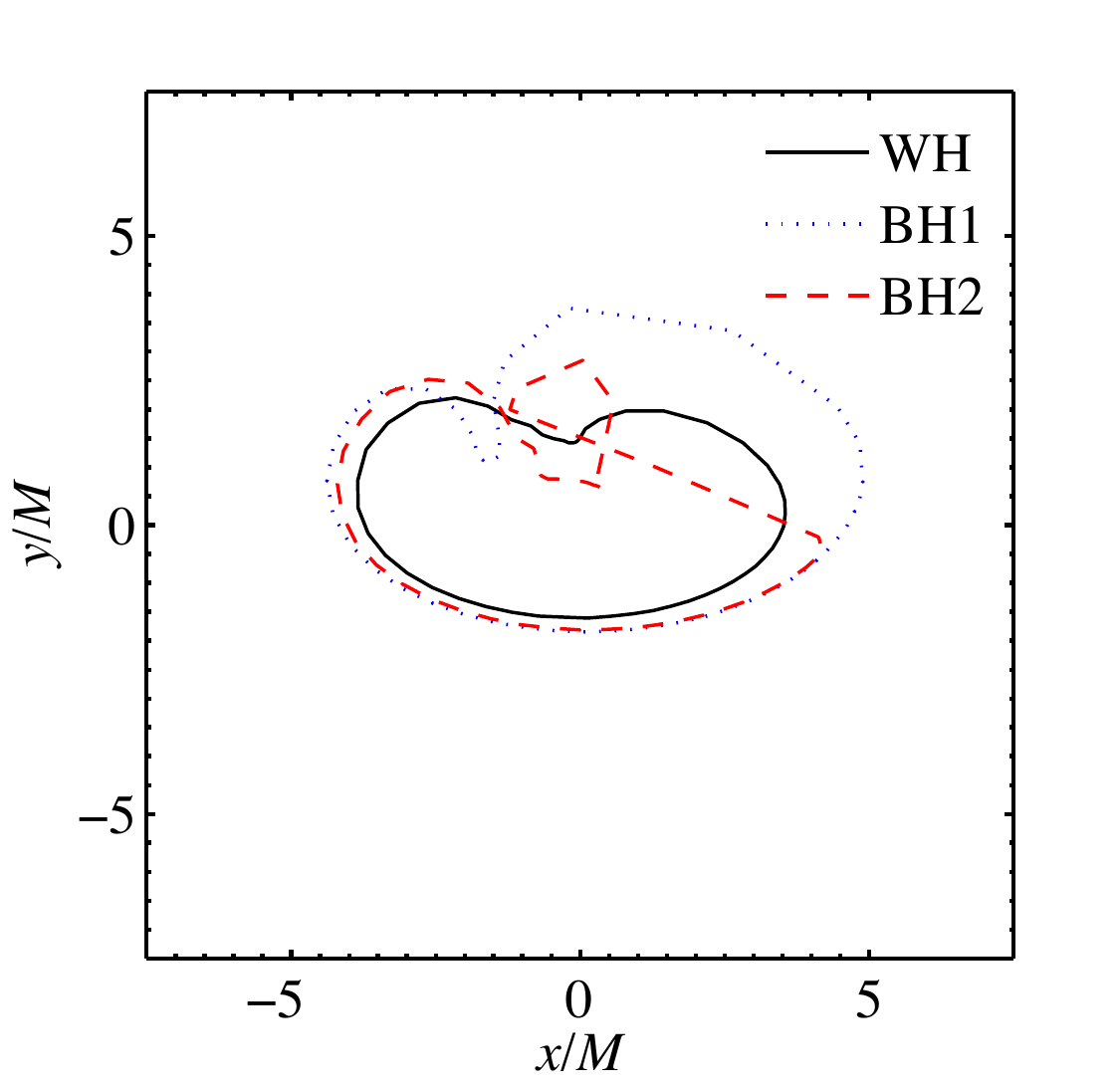} \hspace{0.5cm}
\includegraphics[type=pdf,ext=.pdf,read=.pdf,width=7.5cm]{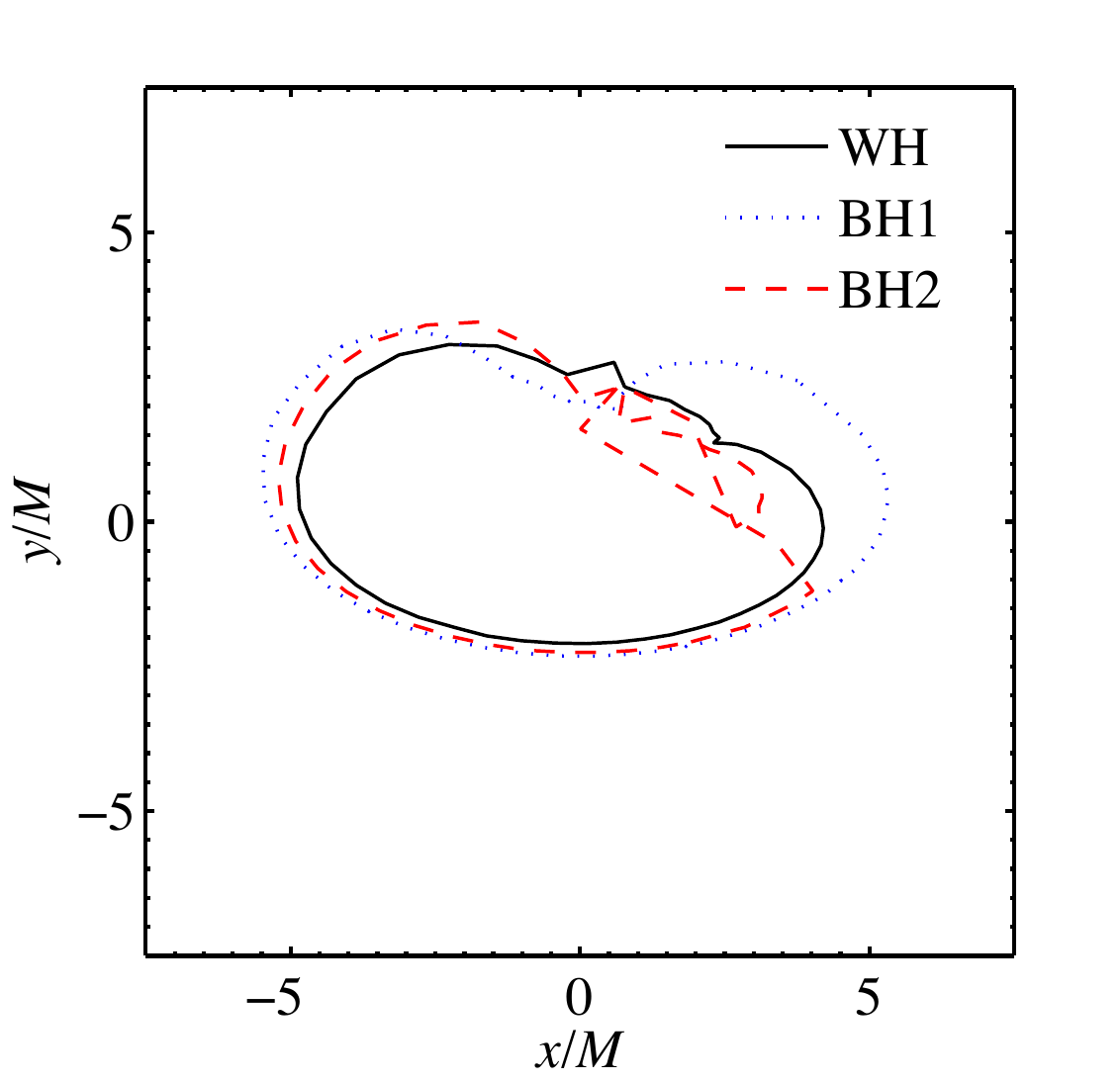}
\end{center}
\caption{Centroid tracks of a hot spot orbiting a WH (black-solid curves), a BH
at the ISCO radius with spin parameter such that the hot spot orbital frequency is the
same as that around the WH (blue-dotted curves), and a BH with spin
parameter $a_*=0.99$ and at the radius with the same Keplerian orbital 
frequency as the one of the hot spots in the other two cases (red-dashed curves).
The orbital radius of the hot spot around the WH is $r_{\rm WH}=2\, M$ (ISCO radius; top 
left panel), $r=_{\rm WH}2.5\, M$ (top right panel), $r_{\rm WH}=3\, M$ (bottom left panel), and 
$r_{\rm WH}=4\, M$ (bottom right panel). The inclination angle of the hot spot orbital plane 
with respect to the observer is $i=60^\circ$ and the hot spot size is $R_{\rm spot} 
= 0.15\, M$. See the text for more details.}
\label{f4}
\end{figure*}

NIR observations may soon be able to directly image hot spots around 
SgrA$^*$. For instance, the VLTI instrument GRAVITY is supposed to be operative
within a few years and be able of astrometric measurements with an angular
accuracy of about 10~$\mu$arcsec and a time resolution around 1~minute.
Such a values have to be compared with the apparent gravitational radius of 
SgrA$^*$, $M$, which is about 5~$\mu$arcsec for an object with a mass of 4~million 
Solar masses and a distance of 8~kpc from us, and with the hot spot orbital period, 
for which current data point out a time scale in the range 13 to 30~minutes 
(presumably due to the different orbital radius in different observations). 
Snapshots of the direct image of hot spots orbiting WHs and BHs are shown in
Figs.~\ref{f2} and \ref{f3}. Every figure compares the three situations mentioned
above, where the images of the hot spot around a WH are in the left column,
the ones of a hot spot around a Kerr BH at the ISCO radius are in the central
column, while the right column is for the images of a hot spot around a Kerr
BH with spin parameter $a_* = 0.99$. Fig.~\ref{f2} shows the cases in which the
hot spot orbiting around the WH has orbital radius $r_{\rm WH}=2\, M$, while Fig.~\ref{f3}
is for the WH hot spot with orbital radius $r_{\rm WH} = 3\, M$. We have thus the two
scenarios already discussed in Fig.~\ref{f1}, and therefore the BH parameters
are the same.

In all these snapshots, the secondary image is dimmer (in some 
snapshots almost absent) and smeared along/near the apparent photon capture 
radius. The apparent photon capture radius, i.e. the one seen by a distant 
observer, was computed in Ref.~\cite{wh2} in the case of the WH in 
Eq.~(\ref{eq-metric}) and it turns out to be about $2.718 \, M$. The angular size 
of the WH on the sky would be about 30~$\mu$arcsec for SgrA$^*$. In the 
case of a Kerr BH, the apparent photon capture radius is about twice that, with a
small dependence on the BH spin and observer's inclination angle. For
SgrA$^*$, it would be around 50~$\mu$arcsec. In particular,
the top panels in Figs.~\ref{f2} and \ref{f3} clearly show the difference between 
the hot spot secondary images in the cases of WHs and BHs. The primary images
are fairly similar, while the secondary images are smeared along the apparent 
photon sphere, which is much smaller in the WH case. Let us also notice that 
such a prediction does not depend on the hot spot model (hot spot size, spectrum, 
observer's viewing angle, etc.), but only on the spacetime metric around the 
compact object. While observationally challenging, the detection of the hot spot 
secondary image and the estimate of the apparent photon capture radius are 
not out of reach in the near future and they seem to be the observational signature 
to distinguish WHs and BHs. Let us also note that the apparent photon 
capture radius depends only on the redshift function $\Phi(r)$, while it is
independent of the shape function $b(r)$~\cite{wh2}.

Lastly, we have computed the hot spot centroid tracks. Fig.~\ref{f4} shows 4 
examples with different hot spot orbital periods. The left panels show the
two cases discussed in Figs.~\ref{f1}-\ref{f3}, in which the hot spot around the WH
has the orbital radius $r_{\rm WH} = 2\,M$ (top left panel in Fig.~\ref{f4}) and $r_{\rm WH} = 3\,M$
(bottom left panel in Fig.~\ref{f4}). In the top right panel in Fig.~\ref{f4}, the
hot spot around the WH has the orbital radius $r_{\rm WH} = 2.5\,M$, while in the bottom
right panel the orbital radius is $r_{\rm WH}=4\,M$. In Fig.~\ref{f4}, the hot spot size is
$R_{\rm spot} = 0.15\,M$ and the observer's viewing angle is still $i = 60^\circ$.
While one may be tempted to argue that the centroid tracks of WHs and BHs
present different features, and therefore that its detection can distinguish
WHs and BHs, as discussed in Ref.~\cite{m4} the exact hot spot model is 
quite important. Since we are considering here a very simple model,
it is not possible to figure out if the detection of the centroid track can be 
used to distinguish WHs and BHs. From the simple model considered here, the 
difference between WHs and BHs does not seem to be so clear and easy to identify.

\section{Summary and conclusions}

WHs are topologically non-trivial structures connecting either two different 
regions of our Universe or two different universes in a Multiverse model. 
While of exotic nature, they are allowed in general relativity and in alternative 
theories of gravity and they are viable candidates to explain the supermassive 
objects harbored at the center of every normal galaxy. In the present paper, 
we have extended the studies of Refs.~\cite{wh1,wh2} and we have further 
investigated if observations can test the possibility that the supermassive 
BH candidates in galactic nuclei are instead WHs. We have focused our 
attention on the specific case of the metric in Eq.~(\ref{eq-metric}), which 
describes an asymptotically-flat non-rotating traversable WH. In Ref.~\cite{wh1},
it was found that such a WH would be consistent with current observations
of the iron K$\alpha$ line detected in the X-ray spectrum of supermassive
BH candidates. In Ref.~\cite{wh2}, it was pointed out that the observation of
the shadow of SgrA$^*$, the supermassive BH candidate at the center of the 
Milky Way, could easily test the possibility that this object is actually a WH 
rather than a BH, because the size of the shadow, which corresponds to 
the apparent photon capture sphere, is much smaller in the WH case than
in the BH one.

In this paper, we have discussed the possibility of testing the presence of a
WH at the center of our Galaxy by observing a hot blob of plasma orbiting
near the ISCO of SgrA$^*$. Such a kind of observations are expected to be
possible soon in the NIR, before the first detection of the shadow of
SgrA$^*$, thanks to the advent of the VLTI instrument GRAVITY. 
We have found that the features of the hot spot secondary image are substantially 
different between a WH and a BH and they probably represent the key-point 
to distinguish the two scenarios. If the hot spot is close to the compact object,
even if it is not necessarily at the ISCO radius, the secondary image shows up
around the apparent photon capture sphere, which is significantly different 
in the two spacetimes. The size of the WH photon capture radius projected 
on the observer sky is indeed about half the BH one and in the case of SgrA$^*$
they correspond, respectively, to about 30 and 50 $\mu$arcsec. The detection
of the direct image of the secondary image of a hot spot could thus test if
SgrA$^*$ is a WH rather than a BH. Such a prediction is very general, in the 
sense that it does not depend on the hot spot model and on the inclination 
angle of the hot spot orbital plane with respect to the line of sight of the
observer. The apparent photon capture radius only depends on the 
spacetime geometry close to the compact object.
Specific features of the secondary image are also encoded in the
hot spot light curve and in its centroid track. However, these features do
depend on the hot spot model and within our simple set-up it is not
possible to figure out if future observations of light curves and centroid tracks
can distinguish WHs and BHs.

A small apparent radius of the photon capture sphere, which is here the true
observational signature to distinguish WHs from BHs, can be found even in other 
contexts. Generally speaking, a similar property can be expected in the case of 
naked singularities, which have also been considered as possible candidates to 
explain the supermassive objects at the centers of galaxies. The absence of an 
event horizon is an indication of the fact that the gravitational field around them is 
weaker than the one around a BH and, at least in some cases studied in the 
literature, the the photon capture sphere of these objects is indeed very 
small~\cite{nks1}. However, these spacetime may be unstable. This is, for 
example, the case of the Kerr spacetime with $a > M$: the existence of an
ergoregion and the absence of event horizons make the spacetime very 
unstable and therefore these objects cannot be considered as serious 
candidates~\cite{nks2}.

Lastly, one may wonder whether the recently announced cloud 
close to SgrA$^*$ and supposed to be soon swallowed by the central supermassive 
objects may be a unique opportunity to test the actual nature of SgrA$^*$~\cite{g2}
and the possible observational predictions in the case of a BH and of a WH. 
Actually, it seems now that the accretion process onto the central object will be 
much slower than what it was initially supposed (just because the gas takes time 
to lose energy and angular momentum) and therefore it is now thought that there 
will be no violent event, and the time scale will be so long that the difference in 
the accretion rate onto SgrA* will probably be irrelevant.


\begin{acknowledgments}
This work was supported by the NSFC grant No.~11305038, 
the Shanghai Municipal Education Commission grant for Innovative 
Programs No.~14ZZ001, the Thousand Young Talents Program, 
and Fudan University.
\end{acknowledgments}


\end{document}